%% file: root.tex
\documentclass[journal]{IEEEtran} 
\usepackage{xcolor}
\usepackage{mathtools,amsfonts}
\usepackage{algorithm}
\usepackage{algpseudocode}
\usepackage{newtxtext}
\usepackage{soul}
\usepackage{array}
\usepackage[T1]{fontenc}
\usepackage[caption=false,font=normalsize,labelfont=sf,textfont=sf]{subfig}
\usepackage{textcomp}
\usepackage{stfloats}
\usepackage{url}
\usepackage{verbatim}
\usepackage{graphicx}
\usepackage{cite}
\usepackage{comment}
\usepackage{amsmath,amssymb, amsthm, enumerate,boxedminipage}
\usepackage{hyperref}
\hypersetup{
    colorlinks=true,
    linkcolor=blue,
    filecolor=magenta,      
    urlcolor=cyan,
    }
\newcommand{\etal}{et al. }

\theoremstyle{definition}
\newtheorem*{prf*}{Proof}
  
\theoremstyle{definition}
\newtheorem*{prfthm*}{Proof of Theorem}
\theoremstyle{definition}
\newtheorem{lem}{Lemma}
\theoremstyle{plain}

\newtheorem{thm}{Theorem}
\newtheorem{prob}{Problem}
\newtheorem{definition}{Definition}
\newtheorem{remark}{Remark}

\begin{document}

%\title{Agent-Based Triangle Counting and its Applications in Anonymous Graphs\thanks{An \textit{Extended Abstract} (2 pages) of this manuscript appeared in the proceedings of AAMAS 2024\cite{aamas_ea}}}

\title{Agent-Based Triangle Counting: Unlocking Truss Decomposition, Triangle Centrality, and Local Clustering Coefficient\thanks{An \textit{Extended Abstract} (2 pages) of this manuscript appeared in the proceedings of AAMAS 2024\cite{aamas_ea}.}}

\author{Prabhat~Kumar~Chand,~\IEEEmembership{Indian Statistical Institute, Kolkata, INDIA, }
        ~Apurba~Das,~\IEEEmembership{BITS Pilani, Hyderabad, INDIA}
        and~Anisur~Rahaman~Molla,~\IEEEmembership{Indian Statistical Institute, Kolkata, INDIA.}
        % <-this % stops a space
}

\markboth{Journal of \LaTeX\ Class Files,~Vol.~14, No.~8, August~2021}%
{Shell \MakeLowercase{\textit{et al.}}: A Sample Article Using IEEEtran.cls for IEEE Journals}


\maketitle

\begin{abstract}
In this paper, we study the problem of \emph{triangle counting} in an arbitrary anonymous graph $G$ with $n$ nodes and $m$ edges using the \emph{mobile-agent model}. Our triangle-counting method serves as a building block for solving related problems such as truss decomposition, triangle centrality, and local clustering coefficient computation. The agents operate synchronously, have distinct identifiers and limited memory, and communicate only when co-located. Starting from an arbitrary placement of $n$ agents, we first obtain a dispersed configuration, elect a leader, construct a spanning tree, and determine the maximum degree $\Delta$ and maximum agent identifier $\lambda$. A BFS tree is constructed separately, which is needed for repeated global communication. Using this setup, the agents enumerate triangles and compute node- and edge-level triangle information, which is subsequently used for truss and centrality computations.  We also complement the theoretical analysis with simulation-based evaluations on representative graph instances. Overall, our results establish a mobile-agent-based framework for these fundamental graph analytics problems in anonymous networks.
\end{abstract}

\begin{IEEEkeywords}
Agent-Based Systems $\cdot$ Autonomous Agents $\cdot$ Distributed Robot Systems $\cdot$ Multi-Robot Systems $\cdot$ Mobile Agents $\cdot$ Triangle Counting $\cdot$  $k-truss$ $\cdot$ Truss Decomposition $\cdot$ Triangle Centrality $\cdot$ Local Clustering Coefficient $\cdot$ Time Complexity $\cdot$ Network Algorithms $\cdot$ Distributed Algorithms 
\end{IEEEkeywords}

\maketitle

\IEEEpeerreviewmaketitle

\input{introduction}

\input{relatedwork}

\input{problem_model}

\input{improvements}

\input{first_algorithm}

\input{k_truss}

\input{application}

\input{simulation}

\input{conclusion}

\bibliographystyle{IEEEtran}
\bibliography{reference}

\end{document}

%% file: introduction.tex
\section{Introduction}\label{sec:intro}

Counting and listing triangles in a graph has received much attention in the last couple of decades as it serves as a building block of complex network analysis~\cite{WS98,B05}. The triangle count in a graph is used for computing the clustering coefficient, one of the most used metrics for network analysis~\cite{WS98,B05}, and triangle centrality~\cite{LB21,AAH22,B2021}. Triangle counting also plays a pivotal role in the hierarchical decomposition of a graph, such as truss decomposition~\cite{WC12} which is an important hierarchical subgraph structure in community detection~\cite{HCQTY14, AZ2017}. Triangle counting is useful in many practical applications. Becchetti \etal~\cite{BBC08} used triangle counts in detecting web spam and estimating the content quality of a web page. Eckmann and Moses~\cite{EM02} have used the clustering coefficient in finding common topics on web pages. Other applications of triangle counting include query optimization in databases~\cite{BKS02}, link prediction in social network~\cite{TDM11}, and community detection in system biology~\cite{JGG14}.

In this paper, we are interested in the triangle counting problem along with its applications in truss decomposition, computing triangle centrality and local clustering coefficient using autonomous agents on anonymous graphs. Let $n$ agents be positioned initially on the nodes (each node has one agent) of an $n$-node anonymous graph $G$. The agents coordinate among themselves to solve the triangle counting problem such that each agent (at a node) outputs (i) a node-based triangle count, (ii) an edge-based triangle count, and the total number of triangles in the graph (see the problem statements in Section~\ref{sec:model}). 

The agent-based model considered in this work (details described in Section~\ref{sec:model}) has been gaining significant attention recently.  For example, there has been some recent work on how to position the agents on nodes of the graph $G$ such that each agent's position collectively forms the maximal independent set (MIS) of $G$~\cite{PramanickSPM23,maximal} or they identify a small dominating set~\cite{run_for_cover} of $G$. Recently, a similar mobile agent model has been used to explore the complexity of constructing a breadth-first search tree~\cite{netys_bfs} and a minimum spanning tree in arbitrary anonymous graphs~\cite{disc_mst,aamas25}. Another related problem is that of \emph{dispersion} in which $k\leq n$ agents have to position themselves at $k$ different nodes of $G$, see~\cite{KS21,spaa25_manish} and the references therein. A solution to the dispersion problem guarantees that $k$ agents are positioned on $k$ different nodes; this is a requirement for the triangle counting problem defined in this paper. Exploration problems on graphs using mobile agents refer to solving a graph analytic task using one or more agents~\cite{D2019}. 

In this work, we consider triangle counting in a simple, undirected, anonymous graph using mobile agents. The motivation comes from scenarios such as private networks in the military or sensor networks in inaccessible terrain where direct access to the network is obstructed, but small battery-powered agents can navigate to learn network structures and their properties for overall network management. The prominent use of agents in network exploration can be seen in areas such as underwater navigation~\cite{CGZ2021}, network-centric warfare in military systems~\cite {LSP2018}, modeling social networks~\cite{ZSW2018}, studying social epidemiology~\cite{ESS2012}, etc. In addition, triangle counting can be effectively applied to drone-based Social Network Analysis (SNA) in scenarios such as drone-assisted event monitoring at public gatherings, protests, or concerts. For example, as demonstrated in~\cite{social1}, drones can identify individuals in close proximity, such as those not wearing masks, and issue social distancing alerts through speakers. In addition, drones can collect data on interactions between individuals, and triangle counting can be used to identify closely knit social groups, discover group dynamics, detect leaders, or monitor potentially suspicious clusters in real time, enhancing situational awareness and crowd management. Furthermore, mobile agents, such as drones, can play a crucial role in Search and Rescue (SAR) operations~\cite{sar1,sar2,sar3,sar4}. In SAR missions, multiple drones can cover large areas, locate missing persons, and maintain communication with the command center. Drones can form resilient triangular communication networks where triangle counting helps identify strong communication links. This ensures robust communication paths between drones and the base station, facilitating quick information sharing. Drones are also being used for Wireless Sensor Network (WSN) deployment in remote or difficult-to-access areas, especially for environmental monitoring~\cite{wsn0,wsn1,wsn2,wsn3}. By identifying dense clusters of sensors or drones, triangle counting enhances network robustness. It helps detect areas of high data redundancy, optimize sensor placement, and improve communication resilience.

\subsection{Our Contributions}\label{sec:our-results}

In this work, we present mobile-agent-based algorithms for triangle enumeration in an $n$-node simple, undirected, anonymous, connected graph $G(V,E)$ with maximum degree $\Delta$, $m$ edges, diameter $D$, and a set of $n$ mobile agents $\mathcal{R}$ with distinct IDs, where $\lambda$ denotes the maximum ID. The agents are initially placed arbitrarily and have no prior knowledge of the graph parameters. A round denotes one complete \emph{Communicate--Compute--Move (CCM)} cycle. Our main contributions are as follows\footnote{Since $\Delta\leq n-1\leq\lambda\leq n^c$, the logarithmic terms involving $\Delta$ and $\lambda$ may be expressed using $\log n$ for comparative analysis with respect to the parameter $n$.}.

\begin{enumerate}
    \item \textbf{Triangle Counting:} Starting from an arbitrary configuration, the agents achieve dispersion, elect a leader $\tilde{r}\in\mathcal{R}$, construct a spanning tree, and compute the parameters $\Delta$ and $\lambda$ in $O(n\log^2 n)$ rounds. Using this setup, the agents compute the number of triangles containing each node and the number of triangles containing each edge in $O(\Delta\log\lambda)$ rounds, and all agents compute the total number of triangles in $O(n+\Delta\log\lambda)$ rounds. For application in truss decomposition, a BFS tree is additionally constructed in $O(D\Delta)$ rounds.

\item \textbf{Applications:}
The agents further use the triangle information to solve three graph analytics problems.
\begin{itemize}
    \item Truss decomposition is computed in $O(m\Delta\log\lambda+mD)$ rounds.
    \item For each node $v\in V$, triangle centrality is computed in $O(\Delta\log\lambda)$ rounds if the total number of triangles $\mathbf{T}(G)$ is known, and in $O(n+\Delta\log\lambda)$ rounds otherwise.
    \item The local clustering coefficient $LCC(v)$ is computed in $O(\Delta\log\lambda)$ rounds.
\end{itemize}
Triangle counting, centrality computation and local clustering coefficient require $O(\Delta\log \lambda)$ bits of memory per agent, whereas truss decomposition requires $O(\Delta^2\log\lambda)$ bits per agent.

\item \textbf{Simulation based Evaluation:}
We also provide simulation-based evaluations on representative graph instances to study the practical behavior of the proposed truss decomposition algorithm. The experiments compare the observed number of truss-upgradation phases with the theoretical worst-case bounds and show that the practical number of phases can be significantly smaller. The code and details can be found online at~\cite{sim_url}.
\end{enumerate}

%% file: relatedwork.tex
\section{Related Work}
\subsection{Triangle Counting}
Triangle counting is a well-studied graph mining problem in both sequential and parallel settings.
\subsubsection{Sequential Algorithms}In 1985, Chiba and Nishizeki~\cite{CN-1985} proposed an algorithm for counting all triangles in a simple graph by computing the intersection of the neighborhoods of the adjacent vertices with time complexity $O(m^{\frac{3}{2}})$ where $m$ is the number of edges in the graph. Other sequential algorithms for triangle counting are based on \textit{vertex-iterator}~\cite{S2007} and \textit{edge-iterator}~\cite{IR1977}. In the \textit{vertex iterator} based technique, it iterates over each vertex $v$ of the graph and intersects the adjacency list of each pair of the neighbors of $v$. In the \textit{edge-iterator} based technique, it iterates over each edge and intersects the adjacency list of its two endpoints.
\subsubsection{Parallel Algorithms}In general, a number of distributed and parallel algorithms for triangle enumeration have been proposed for various models (distributed memory, shared memory, multi-core machines, message passing interface (MPI), etc.). In~\cite{AKM13}, the authors implemented an MPI-based distributed memory parallel algorithm, called PATRIC, for counting triangles in massive networks. Shun \etal in~\cite{ST15} designed multi-core parallel algorithms for exact, as well as approximate, triangle counting and other triangle computations that scale to billions of nodes and edges. In ~\cite{AKM19}, two efficient MPI-based distributed memory parallel algorithms for counting the exact number of triangles in big graphs were presented. They achieved a faster algorithm using overlapping partitioning and efficient load balancing schemes, while a space-efficient one was achieved by dividing the network into non-overlapping partitions. Ghosh \etal in~\cite{GH20} presented a simple MPI-based graph triangle counting
method for shared and distributed-memory systems, which assumes a vertex-based underlying graph distribution called TriC. It was later improved in~\cite{G22}. A detailed account of related works on triangle enumeration for various model set-ups may be found in~\cite{VLPP17,BBC08,BKS02,TDM11,L2008,S2007,GLGBG12,SS11,PSKP2014}. 

\subsection{Truss Decomposition}
In~\cite{C08}, Cohen introduced trusses as a relaxation of cliques and defined a $k$-truss as a non-trivial connected subgraph in which every edge belongs to at least $k-2$ triangles. He also established polynomial-time algorithms for identifying $k$-trusses. Since then, truss decomposition has been studied extensively in several computational models~\cite{HCQTY14,CCC15,HLL16,KYCSZ17}. Existing methods are broadly divided into serialized algorithms for small and medium-sized graphs and parallel algorithms for large graphs. In~\cite{WC12}, Wang \etal proposed an improved in-memory serialized algorithm that computes all $k$-trusses for $k\geq3$ in $O(m^{1.5})$ time using $O(m+n)$ memory, together with two I/O-efficient algorithms for graphs that do not fit in main memory. To handle massive graphs, several parallel approaches were subsequently developed. In~\cite{KM17}, the authors parallelized the algorithm of~\cite{WC12} using data structures suited to concurrent updates. Sariyuce \etal in~\cite{SSP17} applied the iterative h-index method of~\cite{LTQH16} to nucleus decomposition, of which truss decomposition is a special case, and established convergence bounds for synchronous and asynchronous variants. The synchronous version uses a fixed snapshot in each iteration, whereas the asynchronous version immediately uses the latest values. Voegele \etal in~\cite{VLPP17} proposed a parallel graph-centric $k$-truss algorithm and used the fact that every $k$-truss is contained in a $(k-1)$-core to reduce the number of processed edges. Jian Wu \etal in~\cite{WGST18} further engineered the serialized and parallel algorithms of~\cite{WC12} and~\cite{SSP17}, reducing memory usage through optimized data structures and WebGraph. Finally,~\cite{EDST21} studied truss decomposition in probabilistic graphs using h-index updates and derived an upper bound on the number of iterations required for convergence. Most of these algorithms are also evaluated through extensive experiments.

\subsection{Agent-based Computations on Graphs:}~In~\cite{SBN2010}, Sudo \etal considered the exploration problem with a single agent in undirected graphs. Starting from an arbitrary node, the agent must explore all nodes and edges in the graph and return to the starting node. The authors used a whiteboard model that reduced the memory requirement per agent. In \cite{KA2018}, the authors explored the graph exploration problem using Depth-First-Search and studied the trade-off between node memory and agent memory. In~\cite{DDKP2015}, Dereniowski \etal proposed an algorithm for collective graph exploration with a team of $k$ ($k$ being polynomial size) agents in $O(D)$ time. They also obtained near-tight bounds on the asymptotic relationship between exploration time and team size for large $k$ in both the local and the global communication models. Further results on agent-based graph exploration and dispersion (spreading agents across the graph so that there is at most one agent at each node) can be found in several recent papers~\cite{dis03,dis04_cluster,dis08_global,dis06_byz1,dis07_byz2,dis12_fault,dis_fault}.

%% file: problem_model.tex
\section{Model and Problem Definitions}\label{sec:model}

\noindent\textbf{Graph:} The underlying graph $G(V,E)$ is connected, undirected, unweighted and anonymous with $|V| = n$ nodes and $|E| = m$ edges. The nodes of $G$ do not have any distinguishing identifiers or labels. %as well as these nodes are unaware about each other's location. 
The nodes do not possess any memory and hence cannot store any information. The degree of a node $v\in V$ is denoted by $\delta(v)$, and the maximum degree of $G$ is $\Delta$. Edges incident on $v$ are locally labeled using port numbers in the range $[0,\delta(v)-1]$. A single edge connecting two nodes receives independent port numbering at the two ends. The edges of the graph serve as \emph{routes} through which the agents can commute. Any number of agents can travel through an edge at any given time.  \\

\noindent\textbf{Mobile Agents:} We have a collection of $n$ agents $\mathcal{R} = \{r_1,r_2,...,r_n\}$  residing on the nodes of the graph in such a way that each node is occupied by a distinct ID agent at the start (known as \emph{dispersed} configuration in the literature). Each agent has a unique ID (we bound the IDs of the agents by $n^c$, for some large constant $c$, to accommodate the ID space of the agents within an $O(\log n)$ bit ID field) and has limited memory to store information. An agent cannot store the whole graph structure information within its memory. An agent retains its memory as long as needed, and it can be updated as required. Two or more agents can be present (\emph{co-located)} at a node or pass through an edge in $G$. However, an agent is not allowed to stay on an edge. An agent can recognize the port number through which it has entered and exited a node. The agents do not have any visibility beyond their (current) location at a node. An agent at node $v$ can only see the adjacent ports (connecting to edges) at $v$. Only the collocated agents at a node can sense each other and exchange information. As in the standard local-communication model for mobile agents (e.g., \cite{run_for_cover,disc_mst,aamas25}), co-located agents can exchange the contents of their memories in a single round. Thus, unlike the CONGEST model, there is no $O(\log n)$-bit message-size restriction per round. An agent also carries its entire local memory while moving, and the time required for local information transfer between co-located agents is not accounted for in the time complexity analysis, as is standard in such models.

\noindent \textbf{Communication Model:} We consider a synchronous system where the agents are synchronized to a common clock. As mentioned earlier, we consider the local communication model where only co-located agents (i.e., agents at the same node) can communicate among themselves.  \\

\noindent\textbf{Time Cycle:} Each agent $r_i$, on activation, performs a $Communicate-Compute-Move$ $(CCM)$ cycle as follows.
\begin{itemize}
    \item[-]\textbf{Communicate:} $r_i$ may communicate with other agents present at the same node as itself.
    \item[-]\textbf{Compute:} Based on the gathered information and subsequent computations, $r_i$ may perform all manner of computations within the bounds of its memory. 
    \item[-]\textbf{Move:} $r_i$ may move to a neighboring node using the computed exit port. 
\end{itemize}

\noindent An agent can perform the $CCM$ task in one time unit, called {\em round}. The \textbf{time complexity} of an algorithm is the number of rounds required to achieve the goal. The \textbf{memory complexity} is the number of bits required by each agent to execute the algorithm. 

Table~\ref{tab:notations} provides a quick reference of the notations used throughout the paper.

\begin{table}[!ht]
    \centering
    \begin{tabular}{c|c}
        \textbf{Symbols} & \textbf{Meaning}  \\
        \hline
         $G(V,E)$ & Graph with edge-set $E$ and node-set $V$\\
         $n,m$ & Number of nodes and edges of G respectively\\
         $\mathcal{R}$ & Collection of $n$ mobile agents $\{r_1,r_2,\dots,r_n\}$\\
         $\tilde{r}$ & Leader agent\\
         $\lambda$ & ID of the agent with maximum ID\\
         $\delta(v)$ & Degree of the node $v$\\
         $\Delta$ & Highest degree of node in $G$\\
         $h_T$ & Height of the BFS tree of $G$ rooted at $\tilde{r}$\\
         $D$ & Diameter of $G$\\
         $T_k$ & $k-truss$ of $G$\\
         $T(v)$ & Number of triangles with $v$ as vertex (node)\\
         $T(G)$ & Total triangle count of $G$\\
         $N(v)$ & $\{u:(u,v)\in E(G)\}$\\
         $N_{\mathbf{T}}(v)$ & $\{u\in N(v):N(u)\cap N(v)\neq\phi\}$\\ 
         $N^+_{\mathbf{T}}(v)$ & $\{v\}\cup N_{\mathbf{T}}(v)$\\
         $LCC(v)$ & Local Clustering Coefficient of node $v$\\
         \hline
    \end{tabular}
    \caption{Notations used in the paper.}
    \label{tab:notations}
\end{table}

\subsection{Problem Statements}

\begin{prob}
\textbf{Triangle Counting using Mobile Agents: }\label{def: tc_robots}
Consider an undirected, simple, connected anonymous $n$-node graph $G=(V,E)$ and a collection $\mathcal{R}=\{r_1,r_2,\dots,r_n\}$ of $n$ agents, each of which is initially placed distinctly at each node of $G$. We solve the following problems. 

\begin{enumerate}
    \item[(a)] Node-Based Triangle Counting: To count the number of triangles with a given node as a vertex.
    \item[(b)] Edge-Based Triangle Counting: To count the number of triangles based on a given edge.
    \item[(c)] Total Triangle Counting: To count the total number of triangles in the graph $G$.
\end{enumerate}
\end{prob}

\begin{prob}
\textbf{Truss Decomposition, Triangle Centrality and Local Clustering Coefficient: }\label{def: appl_robots}
Consider an undirected, simple, connected anonymous $n$-node graph $G=(V,E)$ and a collection $\mathcal{R}=\{r_1,r_2,\dots,r_n\}$ of $n$ agents, each of which is initially placed distinctly at each node of $G$. The $n$ autonomous agents coordinate among themselves to solve the (i) \emph{Truss Decomposition Problem} and compute (ii) \emph{Triangle Centrality}, and (iii) \emph{Local Clustering Coefficient} of a given node. 
\end{prob}

In this paper, we study the above problems and aim to solve them while minimizing both time and memory-per-agent as much as possible. 
The $n$-agent setting considered in this work corresponds to a full-coverage model, where each node is continuously represented by a mobile computational unit. After dispersion, this can be viewed as a movement-based realization of the local message-passing model, where information is exchanged through coordinated meetings rather than direct communication. This provides a clean abstraction for studying graph-analytic tasks under strict locality constraints, where nodes themselves do not store state or forward messages and all computation is carried by agents. From an algorithmic perspective, the full-coverage model serves as a baseline for understanding the intrinsic cost of local coordination required for problems such as triangle counting and truss decomposition. In particular, it isolates the cost of neighbor discovery and synchronization (e.g., Protocol MYN) without the additional complexity of exploration.
At the same time, this setting naturally supports extensions to fewer-agent deployments. If only $k<n$ agents are available, the same framework can be simulated by allowing agents to represent multiple nodes over time, for example, through phased oscillations, where an agent periodically visits and maintains
states for a small neighborhood, or by placing agents on a dominating set and using them as local representatives for uncovered nodes. Such reductions in the
number of agents potentially introduce a corresponding increase in round complexity due to repeated exploration, delayed meetings, and additional synchronization, and thus reflect a fundamental trade-off between agent density and time. Finally, agent-based computation is also relevant in settings where the computational layer is decoupled from the physical nodes, such as software agents in dynamic networks, drone-assisted sensing, or environments with intermittent or proximity-based communication, where agents carry trusted computation and perform local verification or data aggregation. Thus, the full-coverage model provides both a theoretically clean starting point and a practical basis for studying such constrained distributed computations. A discussion on solving these problems using fewer agents may be found in Section~\ref{sec:conclusion}.

%% file: improvements.tex
\section{Preparation and Setup}\label{sec:improvement}
In our model, before the start of each of the main algorithms, we assumed that each of the $n$ agents is placed on a distinct node of the $n$-node graph, i.e., in a \emph{dispersed} initial configuration, and that every agent has access to certain global parameters, namely $\Delta$ (the maximum degree), $D$ (the diameter) and $\lambda$ (the maximum agent ID). While these assumptions simplify algorithmic design, they are less realistic. In practice, agents may begin in arbitrary, uneven distributions, and global information such as $\Delta,D$ or $\lambda$ is typically unavailable. To address these limitations, we extend our framework to a more general setting. Drawing on recent advances in~\cite{aamas25,spaa25}, we present an algorithm that eliminates both the dispersed-initial-placement requirement and the assumption of prior knowledge of global parameters.

The preparation phase is an initialization step that allows the agents to learn the required parameters without assuming any prior global knowledge. It has two principal steps. In the first step, the agents form a dispersed configuration, with one agent placed at each node, evaluate the values of $\Delta$ and $\lambda$, and construct a spanning tree rooted at an elected leader. This spanning tree is sufficient for global aggregation and dissemination in the basic applications considered in this paper, such as \emph{counting the total number of triangles} and computing \emph{triangle centrality}. For more complex applications such as \emph{truss decomposition}, where global information may need to be propagated repeatedly, we additionally construct and use a BFS structure as a second step. In such cases, the later computation phases dominate the overall running time, making the initialization cost asymptotically insignificant.

Now, the first step is to ensure dispersion from an arbitrary initial configuration. If agents start in clusters, one option is the optimal $O(n)$-round dispersion algorithm of~\cite{spaa25_manish}. However, this algorithm does not detect when dispersion has finished, so the agents cannot know when to begin the actual triangle-counting task. For this reason, we instead rely on the algorithm of~\cite{aamas25}, which takes $O(n \log^2 n)$ rounds but guarantees termination detection. A useful feature of this algorithm is that leader election is performed alongside dispersion, and it does not require any prior knowledge of global parameters. Thus, starting from any arbitrary placement of agents, by applying~\cite{aamas25} we obtain a dispersed configuration with termination detection and a unique leader $\tilde{r}$ among the $n$ agents. This provides exactly the starting point needed for our framework.

Now that we have a dispersed initial configuration, the next step is to determine the parameters $\Delta$ and $\lambda$, which are required in our main algorithm to ensure that each agent systematically meets all of its neighbors. To eliminate the assumption of prior knowledge of $\Delta$, a recent work~\cite{spaa25} proposed scheduling meetings with specific pre-designated neighbors rather than scanning all neighbors at once. This approach avoids unnecessary waiting at low-degree nodes (agents no longer need to wait up to $\Delta$ rounds) and thus removes the need to know $\Delta$ in advance. The method works by padding each agent’s ID with leading zeros to obtain a $\log \lambda$-bit string and then concatenating its bit-wise complement to the left, forming a $2\log \lambda$-bit identifier. The complement ensures that, during traversal, the visiting agent has a bit $1$ where the intended neighbor has a bit $0$, which guarantees their meeting. However, this scheme still relies on knowledge of $\lambda$ (or its upper bound). To overcome this final dependency, the agents first construct a spanning tree, which is then used to compute both $\lambda$ and $\Delta$ without assuming any prior global knowledge. 

\subsubsection*{A Parallel Spanning Tree Construction}\label{sec:parallel_span} Before performing this step, we first obtain a dispersed configuration of the agents and an elected leader, $\tilde{r}$. The construction of the spanning tree is based on~\cite{spaa25}, which we explain briefly here. For this, each agent maintains the following variables: $child$ (the outgoing port numbers through which an agent has moved to undiscovered adjacent nodes and has discovered new agents), $parent$ (the incoming port number through which a particular agent has been first discovered and has been assigned as a $child$), $completion$ (a boolean, initially \texttt{false}, set to \texttt{true} when all children report completion of exploration), and $nextport$ (the port number for the next neighbor to be explored).

The process begins with the leader $\tilde{r}$ exploring its smallest available port, thereby discovering a new agent $ r$. In the next round, both $\tilde{r}$ and $r$ proceed in parallel, each exploring its smallest unvisited port. This simultaneous exploration continues across the system: whenever an agent $r_i$ is discovered by another agent $r_j$, it records the discovery port as its $parent$ and begins exploring its own neighbors, while $r_j$ updates its $child$ list accordingly. Any redundant discovery attempts are discarded, ensuring that each agent is integrated into the tree only once. The parallelism of this strategy lies in the fact that all active agents simultaneously attempt to extend the tree through their $nextport$ values, advancing step by step until every edge incident to the growing tree has been probed. Once an agent $r_i$ exhausts all unexplored neighbors (i.e., sets $r_i.nextport = -1$) and has received $completion$ signals from all its children, it sends a $completion$ message upstream to its $parent$ together with its local degree and the maximum agent ID collected across its sub-tree. These messages propagate upward along parent links until the leader $\tilde{r}$ receives completion reports from all its children, signifying that the entire spanning tree has been successfully constructed and that all global statistics (such as degrees and maximum IDs) have been aggregated. At this stage, $\tilde{r}$ initiates a controlled broadcast to disseminate the collected values back to every agent: in the first round, it transmits the values to its first child, in the next round both $\tilde{r}$ and that child forward the values to their own children, and this doubling-like propagation continues recursively until every agent has been informed. Within $O(n)$ rounds, all agents in the system are guaranteed to receive the disseminated values, completing the spanning tree construction phase and ensuring that every agent holds the same global information. We now construct a BFS tree rooted at leader $\tilde{r}$ and, in the process, collect the height $h_T$ of the BFS tree. At this point, we note the following lemma from~\cite{spaa25}.

\begin{lem}[Spanning Tree~\cite{spaa25}]\label{spanning_tree_time}
    Constructing a spanning tree rooted at the leader $\tilde{r}$ and propagating information between $\tilde{r}$ and any agent takes $O(n)$ rounds.
\end{lem}

\subsection{BFS Tree Construction}\label{sec:the_bfs} Next, we use the collected value of $\Delta$ to construct a BFS tree. Although the spanning tree already provides a communication framework, a BFS tree gives an optimal level-based hierarchy and hence reduces communication overhead. This is useful in algorithms such as \emph{Truss Decomposition} (Section~\ref{sec:truss_decompose}), where agents repeatedly communicate to check progress. For instance, in a complete graph, an arbitrary spanning tree may have height $n-1$, whereas a BFS tree has height one. The BFS tree also allows the agents to compute its height $h_T$, which is later used in the BFS communication procedure (Section~\ref{sec:comm_bfs}) and truss decomposition (Section~\ref{sec:truss_decompose}).

% As described earlier, to ensure the termination of the \emph{Truss Decomposition} algorithm, the $n$ agents must communicate to verify that no further edges are scheduled for upgrading and that no additional iterations are required. This consensus information must be disseminated across all agents to confirm the completion of the algorithm. However, due to the constraints of local communication, propagating such information globally becomes the most time-intensive part of the algorithm. 

% To achieve termination through a global consensus, the agents must repeatedly execute \textsc{Protocol MYN} for at least $D$ rounds, where $D$ is the graph's diameter, each time an edge is flagged for further upgrading. The algorithm terminates only after confirming that all agents reach the state $change=1$, indicating that no further computation is needed. This process intuitively adds a multiplicative factor of at least $O(D\Delta\log\lambda)$ rounds to the time complexity for each iteration (at-most $m$ times in the worst case), where $m$ is the number of edges.

% The complexity of this process can be significantly reduced if, instead of applying \textsc{Protocol MYN} repeatedly, the agents construct a BFS tree rooted at the minimum ID agent, allowing for more efficient communication across the network.

In this subsection, we present an algorithm that constructs a BFS tree for the graph $G$, rooted at the leader agent, denoted as $\tilde{r}$. The primary purpose of this BFS tree is to compute a global value $x = f(x_1, x_2, x_3, \dots, x_n)$ from the individual values $x_i$ provided by each agent $r_i$, and then disseminate the computed value $x$ to all agents in the network in minimum amounts of time. To facilitate the construction of the BFS tree, each agent is equipped with the following variables.

\begin{itemize}
    \item $r_i.parent$, $r_i.child -$ used for storing respective $child$ and $parent$ pointers for $r_i$. We recall that each node $u$ with degree $\delta_u$ has a local port-numbering $=\{0,1,\dots,\delta_u-1\}$. The $r_i.parent$ pointer stores the port number that connects $u$ to its parent agent. Similarly, the $child$ pointers store the port numbers that connect the node $u$ to its children. An agent has at most $\Delta$ child pointers.  Initially, $r_i.parent,r_i.child=\perp$. The child pointers use $O(\Delta\log\Delta)$ bits overall.
    \item $r_i.level -$ denotes the level assigned to an agent relative to the tree that $r_i$ is part of. Initially, $r_i.level=-1$
    \item $r_i.visitor$ - a variable set to $1$, if $r_i$ is a visitor to another node from its own node. When $r_i$ is at its home node, the $visitor$ value is $0$.
    \item $r_i.completion -$ takes a value of $1$, if all its children have been explored. Initially, $r_i.completion=0$
\end{itemize}

The leader agent $\tilde{r}$ begins constructing a BFS tree, with itself as the root. The process proceeds level by level. First, $\tilde{r}$ marks itself as level $0$, initiating the algorithm by visiting all its neighboring agents one by one. As $\tilde{r}$ visits each neighbor, it marks them as its children, and these newly marked children also mark $\tilde{r}$ as their parent simultaneously.

To explore the graph in levels, agents use the variable $level$. When $\tilde{r}$ identifies its children (and they, in turn, recognize $\tilde{r}$ as their parent), these agents update their level to $\tilde{r}.level + 1$. The children of $\tilde{r}$ then begin exploring the next level in search of their own children. However, before they proceed, the children of $\tilde{r}$ at level $1$ must ensure that all their siblings have been explored by $\tilde{r}$. Only after this can agents at the next level begin exploring in parallel. This synchronization is crucial to correctly constructing the BFS tree. If any child prematurely extends the tree before all siblings have been explored, it could result in a tree with unnecessarily long paths, violating the BFS structure. 

Below is an outline of the algorithm for an agent $r_i$ during rounds $2k\Delta$ to $2(k+1)\Delta$ (Algorithm~\ref{alg:bfs}). All references to agents present at a visited node exclude the visiting agent $r_i$ itself.
\begin{enumerate}
    \item \textbf{Case 1}: $r_i.level=k$
    \begin{itemize}
        \item $r_i$ visits its neighbors one by one using increasing port numbers (recall that each node has a local port numbering for every outgoing edge). Before each departure, it sets $r_i.visitor\leftarrow 1$. Let one of its neighbors be $v$.
        
        \item \textbf{Sub-case 1}: If $r_i$ meets a single agent $r_t$ at $v$ with $r_t.visitor=0$ and $r_t.level=-1$ (implying that the resident agent has not been discovered before), it marks $r_t$ as its child, and simultaneously $r_t$ records the port at $v$ through which $r_i$ arrived as its parent port. Additionally, $r_t$ sets $r_t.level\leftarrow r_i.level+1$.
        
        \item \textbf{Sub-case 2}: $r_i$ finds multiple agents at $v$.
        \begin{itemize}
            \item \textbf{1: Original resident present} - If the original resident $r_t$ of $v$ has $r_t.visitor=0$ and $r_t.level=-1$, it accepts the request from the visitor agent with the lowest ID, including $r_i$ among the candidates. If $r_i$ is that agent, it updates its child pointers accordingly. The newly discovered agent records the arrival port of the selected visitor as its parent port and sets its level to $r_i.level+1$. If the resident already has a non-negative level, its parent and level remain unchanged.
            
            \item \textbf{2: Original resident absent} - If the original resident is absent, all agents present are visitors, and no parent-child relationship is established.
        \end{itemize}
        
        \item \textbf{Sub-case 3}: If $r_i$ finds no other agent at $v$, no parent-child relationship is established. This indicates that the resident agent of that particular node is exploring at the same level as $r_i$.
        
        \item \textbf{Sub-case 4}: If $r_i$ encounters a single agent $r_t$ that is already assigned a non-negative level, it leaves the parent and level of $r_t$ unchanged.
        
        \item In every sub-case, including successful discovery, $r_i$ returns to its home node and sets $r_i.visitor\leftarrow 0$ before proceeding to the next neighbor.
        
        \item If all the outgoing ports (neighbors) have been visited, $r_i$ remains at its home node until the phase ends.
    \end{itemize}
    
    \item \textbf{Case 2}: $r_i.level\neq k$: In this case, $r_i$ remains at its home node with $r_i.visitor=0$ during the $2\Delta$ rounds. If $r_i.level=-1$ and it is visited by neighboring agents at level $k$, it accepts the visitor with the lowest ID as its parent, records that visitor's arrival port as its parent port, and sets $r_i.level\leftarrow k+1$. If $r_i$ already has a non-negative level, its parent and level remain unchanged.
\end{enumerate}

\begin{algorithm}[t]
\caption{\small Action of agent $r_i$ during rounds $2k\Delta$ to $2(k+1)\Delta$}
\label{alg:bfs}
\begin{algorithmic}[1]
\Statex \textit{Counts of agents at a visited node exclude $r_i$.}
\If{$r_i.level=k$} \Comment{Case 1}
    \For{each neighbor $v$ of $r_i$ in increasing port order}
        \State $r_i.visitor\gets 1$
        \State Move $r_i$ to $v$
        \If{exactly one other agent $r_t$ is at $v$}
            \If{$r_t.visitor=0$ and $r_t.level=-1$}
                \Comment{Sub-case 1}
                \State $r_i$ marks $r_t$ as child
                \State $r_t.parent\gets$ port at $v$ through which $r_i$ arrived
                \State $r_t.level\gets r_i.level+1$
            \Else \Comment{Sub-case 4}
                \State Leave existing parent and level unchanged
            \EndIf
        \ElsIf{multiple other agents are present at $v$}
            \Comment{Sub-case 2}
            \If{resident $r_t$ with $r_t.visitor=0$ and $r_t.level=-1$ exists}
                \State $chosen\gets$ visitor with lowest ID, including $r_i$
                \If{$r_i=chosen$}
                    \State $r_i$ marks $r_t$ as child
                    \State $r_t.parent\gets$ port at $v$ through which $r_i$ arrived
                    \State $r_t.level\gets r_i.level+1$
                \EndIf
            \EndIf
        \Else \Comment{Sub-case 3: no other agent is present}
            \State No parent-child relationship is established
        \EndIf
        \State Return $r_i$ to its home node
        \State $r_i.visitor\gets 0$
    \EndFor
    \State Remain at home node until the phase ends
\Else \Comment{Case 2}
    \State $r_i.visitor\gets 0$
    \State Remain at home node throughout the phase
    \State On each visit, respond as follows:
    \If{$r_i.level=-1$ and visitors at level $k$ are present}
        \State $r_j\gets$ visitor with lowest ID
        \State Accept $r_j$ as parent and inform $r_j$
        \State $r_i.parent\gets$ port through which $r_j$ arrived
        \State $r_i.level\gets k+1$
    \EndIf
\EndIf
\end{algorithmic}
\end{algorithm}

To complete the BFS tree construction, the exploration proceeds in phases of length $2\Delta$, including a final exploration phase for agents at level $h_T$, where $h_T$ denotes the BFS tree's height. Since the agents do not initially know the value of $h_T$, a termination mechanism---similar to that used in the spanning tree construction---is employed. An agent $r$ is ready to report completion only after it has explored all its incident ports and received completion messages from all its children. A leaf becomes ready after exploring all its incident ports. Each non-root agent then sends a completion message to its parent together with the maximum of its own level and the level values reported by its children. The algorithm terminates once the leader (root of the BFS tree) has explored all its incident ports and received completion information from all its children, thereby finalizing the BFS tree. The root then disseminates the maximum level $h_T$ of the constructed BFS tree to all agents via the established BFS structure. From the discussion above, we have the following lemma.

\begin{lem}[BFS Tree Construction]\label{spanning_tree}
    The $n$ agents construct a BFS tree for $G$ rooted at $\tilde{r}$ in $O(h_T\Delta)$ rounds.  
\end{lem}

Although the BFS tree has been constructed, one challenge remains: the agents must establish an efficient procedure for propagating values across the tree. Many algorithms, such as the \emph{Truss Decomposition} described later, require repeated communication over the spanning tree to achieve global agreement on certain values. Thus, designing a mechanism that enables faster and more reliable information transfer is crucial. While the BFS tree provides the necessary topological structure, we now describe the actual communication mechanism.

\subsection{Communication via the BFS Tree}
\label{sec:comm_bfs}

The BFS tree constructed during the setup phase serves as the communication backbone among the agents. Since each agent knows its parent and the height $h_T$ of the tree, any value can be propagated to the leader $\tilde{r}$ by forwarding it upward through the parent pointers until it reaches the root. Broadcasting a value from $\tilde{r}$ is performed in a parallel manner to avoid a sequential $O(n)$ process. Whenever a broadcast is required, each child alternates between its own node and its parent over the next $2h_T$ rounds. In the first two rounds, the children of $\tilde{r}$ receive the value and return; in the next two rounds, the grandchildren receive it, and the process continues level by level. Thus, after two rounds per level, every agent receives the value, and the entire broadcast completes in $O(h_T)$ rounds. This allows efficient upward aggregation and downward dissemination of information using the BFS-tree structure. The pseudo-code is provided in Algorithm~\ref{alg:bfs_comm}

\begin{algorithm}[t]
\caption{Communicating via the BFS Tree}\label{alg:bfs_comm}
\begin{algorithmic}[1]
\Procedure{PropagateToLeader}{$value$}
    \State All agents start at their home nodes
    \State Each agent initializes its aggregate to its local $value$
    \For{$\ell=h_T$ down to $1$}
        \State Agents at levels other than $\ell$ remain at home
        \State Each agent at level $\ell$ moves to its parent's home node
        \State Each parent combines the received aggregates with its own
        \State Visiting agents return to their home nodes
        \Comment{Two rounds per level}
    \EndFor
    \State Leader $\tilde{r}$ holds the global aggregate
\EndProcedure

\vspace{0.5em}

\Procedure{BroadcastFromLeader}{$value$}
    \State All agents start at their home nodes
    \State Leader $\tilde{r}$ holds $value$ and remains at home
    \State Mark all other agents as uninformed for this broadcast
    \For{round $r=1$ to $2h_T$}
        \ForAll{non-root agents $r_i$ with parent $p_i$ in parallel}
            \If{$r$ is odd}
                \If{$r_i$ is uninformed}
                    \State $r_i$ moves to the home node of $p_i$
                \EndIf
            \Else
                \If{$r_i$ is away from its home node}
                    \If{$p_i$ is present and informed}
                        \State $r_i$ receives $value$ and becomes informed
                    \EndIf
                    \State $r_i$ returns to its home node
                \EndIf
            \EndIf
        \EndFor
    \EndFor
    \State All agents hold $value$ and are at their home nodes
\EndProcedure
\end{algorithmic}
\end{algorithm}

\begin{lem}[BFS Tree Communication]\label{spanning_tree_communication}
    The $n$ agents can communicate via the BFS tree in $O(h_T)$ rounds.  
\end{lem}

Since the BFS tree has been created, the leader can take responsibility for computing a certain global functional output based on the input received from each agent locally. We emphasize this in the following lemma.
\begin{lem}\label{output_transmission}
    Each agent can learn the value of the function $f(x_1, x_2, \dots, x_n)$, which is computed from $n$ inputs provided by the $n$ agents, in $O(h_T)$ rounds by utilizing the constructed BFS tree as the framework. 
\end{lem}

\begin{remark}\label{rem:D-h_t}
Let $D$ denote the diameter of $G$, and let $h_T$ denote the height of the BFS tree rooted at $\tilde{r}$. Since $\frac{D}{2}\leq h_T\leq D$, we have $O(h_T)=O(D)$. Hence, the algorithms use the actual BFS-tree height $h_T$, whereas the final complexity bounds are stated in terms of the standard graph parameter $D$.
\end{remark}

We summarize the results of the section below.

\begin{thm}[Initialization]\label{thm:gen_config}
Let $G$ be an $n$-node arbitrary, simple, connected, anonymous graph with maximum degree $\Delta$ and diameter $D$. Let $n$ mobile agents with distinct IDs be initially placed at arbitrary locations in $G$. Then the following holds.
\begin{enumerate}
\item The agents can reach a \emph{dispersed} configuration, with exactly one agent placed at each node of $G$, elect a leader agent denoted by $\tilde{r}$, and construct a spanning tree rooted at $\tilde{r}$ in $O(n\log^2 n)$ rounds. Using this spanning tree, the agents can compute global values such as $\Delta$ and $\lambda$ in an additional $O(n)$ rounds; thereby, the overall initialization cost for this part is $O(n\log^2n+n)=O(n\log^2n)$

\item Further, the leader $\tilde{r}$ can construct a BFS tree rooted at $\tilde{r}$ in $O(n\log^2 n+h_T\Delta)=O(n\log^2 n+D\Delta)$ rounds. Once this BFS tree is constructed, any global communication among the $n$ agents can be performed along the BFS tree in $O(h_T)=O(D)$ rounds, where $h_T$ is the height of the constructed BFS tree.

\end{enumerate}
For this process, an agent uses $O(\log\lambda)$ bits of memory. 
\end{thm}

\begin{remark}
\label{rem:spanning_vs_bfs}
The spanning tree constructed in Lemma~\ref{spanning_tree_time} is sufficient for global communication among all agents. Once this tree is available, any value held by the agents can be aggregated at the leader and then disseminated back to all agents in $O(n)$ rounds. Therefore, for problems such as \emph{counting the total number of triangles} and computing the \emph{triangle centrality} of a node, where only a constant number of global aggregation and dissemination steps are required, we use this spanning tree; this keeps the initialization cost bounded by $O(n\log^2 n)$ rounds. In contrast, although a BFS tree provides a smaller-height structure for global propagation, its construction requires $O(D\Delta)$ rounds, which may be as large as $O(n^2)$ in arbitrary graphs. Hence, we construct and use the BFS tree only for \emph{truss decomposition}, where global information may need to be propagated repeatedly over up to $m$ iterations in the worst case, and the smaller BFS height $h_T$ can substantially reduce the overall communication cost, especially when $D\ll n$.
\end{remark}

 Throughout this work, for the basic applications, we assume that the agents start from a dispersed configuration and have prior knowledge of the required global parameters, such as $\Delta$ and $\lambda$. This assumption is justified since such a state can be reached from any arbitrary initial configuration within $O(n\log^2 n)$ rounds using only $O(\log n)$ bits per agent, even when the agents begin without any knowledge of the network. For applications that require repeated global propagation, such as \emph{truss decomposition}, the BFS tree can be constructed in $O(n\log^2 n+D\Delta)$ rounds, enabling subsequent communication in $O(D)$ rounds.

%% file: first_algorithm.tex
\section{Triangle Counting via Mobile Agents}\label{sec: traingle_counting}
In this section, we develop algorithms for $n$ mobile agents that are initially assumed to be dispersed among the $n$ nodes of the graph $G$ to enumerate the number of triangles in $G$. Since the nodes themselves are memory-less and indistinguishable, the algorithm relies on the memory and IDs of the mobile agents that reside on the nodes of the graph. Also, since the agents cannot communicate among themselves (unless they are at the same node), synchronizing the movement of the agents is another challenge.  

In our algorithm, the agents (technically representing the nodes they are sitting at) first scan their neighborhood. Once all the information about the neighborhood is collected, the agents then count the number of common neighborhoods between two adjacent agents. After each agent receives that count, the sum of each such count stored at each agent is evaluated, which, when divided by six, gives us the number of triangles in $G$. The algorithm runs in three phases. In the first phase, the agents learn their neighbors. In the second phase, the agents check the number of common neighbors with each of their adjacent agents. In this phase, each agent $r_i$ also counts the number of local triangles with $r_i$ as a vertex and the number of triangles with $(r_i,r_j)$ as an edge, where $r_j$ is an adjacent agent to $r_i$. In the third and last phase, each agent collects the local triangle count from every other agent and counts the number of triangles in $G$. We now explain each phase in detail below.

\subsection{Phase 1: Neighborhood Discovery (\textbf{\textsc{Protocol MYN}})}\label{sec: KYN}
Each node is occupied by a distinct (via their IDs) agent before the start of the algorithm, and we represent each node of $G$ by its stationed agent. The algorithm starts with \textit{Phase 1}, where each agent discovers and records its neighbors. 
% As the robots do not have global communication (they can only talk to the robots at the same node) and the nodes themselves are indistinguishable, the robots need to move from node to node to gather information and execute any algorithm. 
As the IDs of the agents are unknown, agents cannot synchronously start scanning the neighborhood. 
With each agent executing the same algorithm, agents may not find other agents at their exact place when they move. Therefore, we need to ensure that each agent gets to record the correct neighbor set of itself. To do this, we exploit the ID bits of the agents. We use the fact that the IDs of the agents are distinct. To this end, we state the following lemma.
\begin{lem}\label{lemma: differnt_bit}
    Let $r_i$ and $r_j$ be two distinct agents in $\mathcal{R}$, with their ID fields being $r_i.ID$ and $r_j.ID$ respectively. Then, there exists at least one dissimilar bit in $r_i.ID$ and $r_j.ID$ with one being $0$ and the other being $1$.
\end{lem}
Let $\lambda$ denote the largest ID among all the $n$ agents. Therefore, the agents use a $\log \lambda$ bit field to store the IDs. Now, to list the neighboring agents, an agent $r_i$ stationed at a node $u$ does the following. Here, $\delta(u)$ denotes the degree of a node $u$ and $\Delta$ denotes the highest degree of a node in $G$. 

\begin{enumerate}
\item For $\log\lambda$ rounds $r_i$ executes the following.
\begin{enumerate}
    \item $r_i$ checks the \textbf{current ID bit} in its ID from the right. (At the start, the \textbf{current ID bit} is the rightmost bit in the ID field).
    \item In the next $2\Delta$ rounds, $r_i$ chooses to do one of the following two:
    \begin{itemize}
        \item If the \textbf{current ID bit} is $0$, $r_i$ waits at its own node for $2\Delta$ rounds.
        \item If the \textbf{current ID bit} is $1$, $r_i$ visits each of its neighbor and back using \textit{port number} $0$ till \textit{port} $\delta(u)-1$. When $r_i$ meets a new agent $r_k$, it checks $r_k$'s \textbf{current ID bit}. If the \textbf{current ID bit} of $r_k$ is $0$, it adds $r_k$ and records it to $r_i$'s neighbors list. (This is to ensure that $r_k$ is the original neighbor of $r_i$ and $r_k$ is not an exploratory agent from a different neighboring node). The agent $r_k$ also simultaneously records $r_i$ as its neighbor as well. The agents ignore any agent that has already been registered.
    \end{itemize}
    \item After the $2\Delta$ rounds have elapsed, the next left bit in the ID field becomes the \textbf{current ID bit}. If no more bits are remaining and $\log\lambda$ rounds have not been completed (implying that $r_i$ has a smaller ID length than $\log\lambda$ bits), $r_i$ assumes the current bit as $0$ and stays back at its own node for the rest of the algorithm.  
\end{enumerate}
\end{enumerate}
After the completion of Phase 1 of our algorithm, it is guaranteed that each agent correctly records all its neighboring agents in $O(\Delta\log\lambda)$ rounds, which we prove in the following lemma.
\begin{lem}\label{lemma: list_neighbours}
   Each agent $r_i$ correctly records the exhaustive list of its neighbor agents in $O(\Delta\log\lambda)$ rounds after the completion of Phase 1.
\end{lem}
\begin{proof}
    Let $r_j$ (placed at node $v$) be a neighbor of $r_i$ (at node $u$). Then there exists a port joining $u$ to $v$ having a port number between $[0,\delta(u)-1]$. Since $r_i$ and $r_j$ have distinct IDs, there exists a bit (say the $p^{th}$ bit from the right) in $r_i.ID$ and $r_j.ID$, which are different from one another (one bit being $0$, the other being $1$)[Lemma~\ref{lemma: differnt_bit}]. Without the loss of generality, let us assume that the $p^{th}$ bit of $r_i$ is $1$ and the $p^{th}$ bit of $r_j$ is $0$. Therefore, when the $p^{th}$ bit becomes the \textbf{current ID bit}, $r_i$ starts exploring its neighbors one by one, during which it finds the agent $r_j$ now stationary at $v$ with \textbf{current ID bit} $0$.  Therefore $r_i$ records $r_j$ as its neighbor (simultaneously, $r_j$ also registers $r_i$ in its neighbor list). In a similar way, $r_i$ records all of its neighbors exhaustively either by visiting a stationary agent in its neighborhood or by meeting an exploratory agent from its neighbor. Phase 1 runs for $\log\lambda$ rounds, with each round consisting of $O(\Delta)$ sub-rounds to allow the agents to visit each neighbor. Therefore, Phase 1 completes in  $O(\Delta\log\lambda)$ rounds.
\end{proof}

Therefore, with the end of Phase 1, each agent has now enlisted its neighboring agents in its memory. We call this particular step \textsc{Protocol MYN}, a protocol for \textit{Meeting-Your-Neighbor}. The algorithm now moves to Phase 2, where each agent $r_i$ counts the number of triangles with $r_i$ as one of its vertices (node).

\subsection{Phase 2: Local Triangle Counting}
In Phase 2, each agent, now equipped with the list of its neighboring agents, visits each of its neighbors once again, in exactly the same manner as described in Phase 1. Whenever the agent $r_i$ meets its neighbor $r_j$, they communicate to find out the number of common neighbors they both have. Since the nodes are anonymous, they are identified using the mobile agents that reside on the nodes. This communication is also used to update the following variables of $r_i$:
\begin{enumerate}
    \item $r_i.nbd\_list:$ A variable which stores the list of all neighbors of agent $r_i$. The size of $nbd\_list$ is $\Delta\log n$ bits as it may store up to $\Delta$ entries in the worst case. 
    \item $r_i.edge(r_j):$ A variable which stores the count of the number of common neighbors of $r_i$ and $r_j$, which represents the number of triangles based on the edge containing $(r_i,r_j)$. $r_i$ has $\Delta$ such variables $r_i.edge(x)$, where $x$ represents another agent that has an edge with $r_i$. This, therefore, requires an additional $O(\Delta\log n)$ memory per agent, which forms the dominant factor in the memory requirement. Each of these variables is initially set to $0$. A variable $r_i.edge(x)$ remains $0$, if there are no triangles with $(r_i,x)$ as an edge.  To correctly compute these values, the agents use the variable $nbd\_list$. 
    \item $r_i.local\_sum:$ Initially set to $0$, adds up the counts of the common neighbors for each distinct neighbor of $r_i$. As $r_i$ finds its neighbors one by one, it cumulatively adds up the count of the number of common neighbors with each of its neighbors. 
    Mathematically, $r_i.local\_sum=\Sigma_j r_i.edge(r_j)$, where the sum runs over every neighboring agent $r_j$ of $r_i$. 
\end{enumerate}
In the given window of $\log\lambda$ rounds (each round containing a $\Delta$ sub-round), as the agents communicate with their neighbors, the variables $r_i.edge(\cdot)$ are updated. At the end of this phase, $r_i$ also builds on the count of the variable  $r_i.local\_sum=\Sigma_j r_i.edge(r_j)$. To get the correct number of triangles with $r_i$ as one of the vertices (local triangle counting), we divide $r_i.local\_sum$ by $2$. We state the reason for the same in the following lemma.
\begin{lem}\label{lemma: local_triangle_counting}
    The number of triangles with agent $r_i$ as a vertex is given by $\frac{1}{2}(r_i.local\_sum)$.
\end{lem}
\begin{proof}
    First, we observe that if the agents $r_i$ and $r_j$ have a common neighbor $r_k$, then $(r_i,r_j,r_k)$ form a triangle with $r_i$ as one of the vertices. Now, the triangle  $(r_i,r_j,r_k)$  has been tallied once while counting the common neighbor of $r_i$ with $r_j$ and re-counted again as a common neighbor of $r_i$ with $r_k$. Therefore, each $(r_i,r_j,r_k)$ with $r_i$ as vertex, is counted twice, once through $r_j$ and the other time through $r_k$. So, the number of triangles with agent $r_i$ as a vertex is exactly half of  $r_i.local\_sum$. 
\end{proof}

Thus, at the end of Phase 2, each agent $r_i$ gets a local count of the number of triangles with $r_i$ as a vertex and a list of the number of triangles that are based on the edges adjacent to $r_i$. In the next phase, the algorithm determines the number of triangles of $G$ using the counts generated in this phase.

\subsection{Phase 3: Counting the Total Number of Triangles}
To find the number of triangles in $G$, we need to take into account the local triangle counts of each agent. First, we need to accumulate the $r_i.local\_sum$ values of each of the $n$ agents and calculate the triangles of $G$ from there. To aggregate the $r_i.local\_sum$ values from each of the $n$ agents, we use the constructed spanning tree as described in Section~\ref{sec:parallel_span}. The agents aggregate their $local\_sum$ values along the spanning tree. The root then broadcasts only the aggregate $S=\sum_{i=1}^{n}r_i.local\_sum$ to all agents. These operations take $O(n)$ rounds (Lemma~\ref{spanning_tree_time}).

\begin{lem}
By the end of Phase 3, which takes an additional $O(n)$ rounds, each agent $r_i$ knows the aggregate $S=\sum_{j=1}^{n} r_j.local\_sum$ from all the n agents. 
\end{lem}

\begin{lem}\label{global_triangle_count}
    The number of triangles in graph $G$ is given by $\frac{1}{6}\Sigma_i (r_i.local\_sum)$, where the sum runs over all the $n$ agents of $G$. 
\end{lem}
\begin{proof}
    The number of triangles with agent $r_i$ as a vertex is given by $\frac{1}{2} r_i.local\_sum$. Now, since each triangle is counted once for every vertex it has, the total number of triangles in $G$ is given by $\frac{1}{3}\Sigma_i \frac{1}{2}(r_i.local\_sum)$.
\end{proof}

\textbf{Notation: } We shall denote the number of triangles containing the vertex (node) $v$ and the total number of triangles in $G$ with $\mathbf{T}(v)$ and $\mathbf{T}(G)$, respectively. For a node $v$ with an agent $r_i$ on it, we use $r_i$ and $v$ interchangeably to denote the node.

\begin{remark}
The variables stored by an agent may depend on different parameters. An agent ID requires $O(\log\lambda)$ bits, an edge identified by two agent IDs requires $O(\log\lambda)$ bits, and a port number requires $O(\log\Delta)$ bits. Since $\lambda<n^c$ for a constant $c$ and $\Delta\leq n-1$, $O(\log\Delta)$ can be expressed as $O(\log \lambda)$.
\end{remark}

At the end of this phase, each agent has computed the number of triangles in $G$, i.e., $\mathbf{T}(G)$. In the following theorem, we summarize the results obtained during the three phases.

\begin{thm}
Let $G$ be an $n$-node arbitrary, simple, connected graph with maximum degree $\Delta$. Suppose $n$ mobile agents, each with a distinct ID from the range $[0,n^c]$ for some constant $c$, are initially placed arbitrarily in $G$, without any prior global knowledge. Let $\lambda$ denote the maximum agent ID. Then:
    \begin{enumerate}
        \item The agents can first disperse across the graph, construct a spanning tree from a designated root, and collect the necessary parameters in $O(n \log^2 n)$ rounds. (Initialization Theorem - Theorem~\ref{thm:gen_config}) (Part $1$). After this initialization step,
        \item Each agent $r_i$ can compute the number of triangles containing $r_i$, i.e., $\mathbf{T}(r_i)$, in $O(\Delta \log \lambda)$ rounds.
        \item Each agent $r_i$ can compute the number of triangles associated with each of its incident edges in $O(\Delta \log \lambda)$ rounds.
        \item Each agent $r_i$ can compute the total number of triangles in $G$, i.e., $\mathbf{T}(G)$, in $O(n+\Delta \log \lambda)$ rounds.
    \end{enumerate}
These computations require $O(\Delta \log \lambda)$ bits per agent.
\end{thm}

%% file: k_truss.tex
\section{Applications: Truss Decomposition, Triangle Centrality and Local Clustering Coefficient}\label{sec:applications}

In this section, we show some applications of our triangle-counting algorithm via mobile agents in an anonymous graph, namely in solving the \emph{truss decomposition} problem and computing the \emph{triangle centrality} metric and \emph{local clustering coefficient} for each node of the graph. Truss decomposition and clustering coefficient calculations using mobile agents offer significant advantages in environments where decentralized and adaptive operations are crucial. In scenarios with limited or unreliable communication infrastructure, mobile agents can perform local computations, reducing the need for extensive data transmission. This capability is especially valuable in temporary or ad hoc networks, such as those formed during disaster response, where agents can quickly analyze the network's structure to identify robust sub-graphs or highly connected areas, ensuring that resources are efficiently allocated. For instance, drones operating in disaster zones can perform truss decomposition to pinpoint areas of strong connectivity, directing rescue efforts where the network of survivors or critical infrastructure is most resilient. In urban environments, mobile agents representing vehicles can compute clustering coefficients to identify highly interconnected regions of a city, enabling more effective traffic management and resource distribution. In the military context, UAVs can employ triangle counting to reveal strategic communication hubs, improving surveillance and ensuring critical areas are secured against potential disruptions.

In the mobile-agent framework, the graph need not be available at one central machine, nor do the nodes need to store information or actively participate in the computation. The agents carry the computation with them: they move through the graph (from one node to another), collect local neighborhood information, and coordinate through controlled meetings. This is useful in passive, damaged, unreliable, intermittently connected, or partially trusted networks, where modifying the nodes or relying on continuous centralized coordination may not be feasible. Moreover, if truss computation is needed only in a selected region or subgraph, agents can be deployed or moved specifically to that part of the network, possibly from external sources, without restarting the computation globally. This portability makes the model suitable for drone-assisted sensing, underwater monitoring, emergency deployments, mobile software-agent networks, and ad hoc systems. In such settings, triangle counts and edge supports provide local redundancy information, while truss decomposition helps identify structurally robust and strongly connected regions, such as reliable relay structures or weakly attached parts of the graph. Thus, the mobile-agent approach supports localized, flexible, and infrastructure-light robustness analysis under local communication and limited memory.

\subsection{Truss Decomposition}
\label{sec:truss_decompose}

We use the triangle-counting techniques from Section~\ref{sec: traingle_counting} to identify $k$-truss subgraphs of $G$ using mobile agents. For $k\geq 2$, a subgraph $T_k$ of $G$ is a $k$-truss if every edge of $T_k$ participates in at least $k-2$ triangles within $T_k$. Thus, the problem of finding $k$-trusses is usually handled through \emph{truss decomposition}, where each edge is assigned a \emph{trussness} value, and the $k$-truss is obtained from all edges whose trussness is at least $k$. 

\begin{definition}[Support]
For a graph $G=(V,E)$, the \emph{support} of an edge $e\in E$ is the number of triangles in $G$ containing $e$.
\end{definition}

\begin{definition}[$k$-truss]
For $k\geq 2$, a $k$-truss, denoted by $T_k$, is the largest subgraph of $G$ in which every edge has support at least $k-2$ with respect to $T_k$. If no such non-empty subgraph exists, then the $k$-truss of $G$ does not exist.
\end{definition}

\begin{definition}[trussness]
The \emph{trussness} of an edge $e$ is the maximum value $k\geq 2$ such that $e$ belongs to $T_k$ but not to $T_{k+1}$.
\end{definition}

Thus, an edge with support zero has trussness $2$. Such an edge does not belong to any nontrivial $k$-truss with $k\geq 3$.

\noindent
\textbf{Truss decomposition algorithm.}
We compute the trussness of every edge using the parallel truss decomposition framework of~\cite{WGST18}. Once the trussness values are known, the edges are partitioned into equivalence classes according to their trussness values. Let $\mathbf{t_i}$ denote the set of edges with trussness $i$, and let $t_{\max}$ be the maximum trussness of any edge in $G$. Then, for any $k$, the corresponding $k$-truss is obtained as $T_k=\bigcup_{i=k}^{t_{\max}}\mathbf{t_i}$. Hence, computing all edge trussness values is equivalent to computing the full truss decomposition of $G$.

Serial truss decomposition first computes the support of every edge and repeatedly removes an edge with minimum support. If an edge $e$ is removed with support value $s$, then its trussness is set to $s+2$. After removing $e$, the supports of other edges appearing in triangles with $e$ are updated, and the process continues. Since the edges must be repeatedly sorted by support, this approach is inherently sequential.

The parallel algorithm of~\cite{WGST18} avoids this sequential sorting by using $h$-index updates. For a set of integers $S$, the $h$-index is the largest integer $h$ such that at least $h$ elements of $S$ are at least $h$. Initially, each edge $e$ is assigned $h(e)=support(e)$. For every triangle containing $e$, say with other two edges $e'$ and $e''$, the value $\min\{h(e'),h(e'')\}$ is inserted into a multi-set $L$. The value of $h(e)$ is then updated using the $h$-index of $L$. This process is repeated in parallel until no $h$ value changes. Finally, the trussness of each edge $e$ is set to $h(e)+2$.

\begin{algorithm}[!t]
\caption{Parallel $K$-Truss Decomposition~\cite{WGST18}}
\label{alg:k-truss-parallel}
\begin{algorithmic}[1]\small
\Function{$k$-Truss-Parallel}{$G, support$}
    \For{each edge $e \in E$}
        \State $h[e] \gets support[e], \; scheduled[e] \gets \text{TRUE}$
    \EndFor
    \State $updated \gets \text{TRUE}$
    \While{$updated$}
        \State $updated \gets \text{FALSE}$
        \For{each edge $e \in E$ in parallel}
            \If{$scheduled[e] = \text{FALSE}$}
                \State \textbf{continue}
            \EndIf
            \State $L \gets \emptyset,\; N \gets \emptyset$
            \For{each triangle containing $e$}
                \State $e',e'' \gets$ the other two edges of the triangle
                \State $N.add(e'),\; N.add(e'')$
                \State $L.add(\min\{h[e'],h[e'']\})$
            \EndFor
            \State $H \gets h$-index of $L$
            \If{$h[e] \neq H$}
                \State $updated \gets \text{TRUE}$
                \For{each edge $e_N \in N$}
                    \If{$H < h[e_N] \leq h[e]$}
                        \State $scheduled[e_N] \gets \text{TRUE}$
                    \EndIf
                \EndFor
                \State $h[e] \gets H$
            \EndIf
            \State $scheduled[e] \gets \text{FALSE}$
        \EndFor
    \EndWhile
    \For{each edge $e \in E$}
        \State $trussness[e] \gets h[e]+2$
    \EndFor
    \State \Return $trussness$
\EndFunction
\end{algorithmic}
\end{algorithm}

\begin{figure}[!t]
    \centering
    \includegraphics[width=1\linewidth]{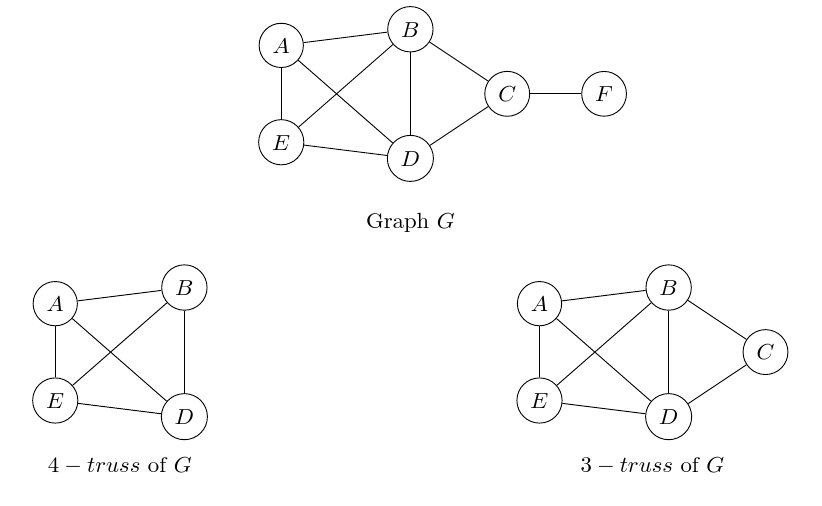}
    \caption{Illustration of the $4$-truss and $3$-truss of graph $G$ with $6$ nodes and $9$ edges. In a $k$-truss, every edge participates in at least $k-2$ triangles.}
    \label{fig:truss_dec}
\end{figure}

Before describing the mobile agent adaptation, we first illustrate the execution of the parallel algorithm using the example in Fig.~\ref{fig:truss_dec}. For the graph $G$, the edge set is $E=\{AB,AD,AE,BC,BD,BE,CD,CF,DE\}$. Let $h(e)$ denote the current $h$-value of an edge $e$, initially set to its support, and let $scheduled(e)$ indicate whether $e$ is marked for evaluation in the next iteration. The variable $updated$ records whether the $h(\cdot)$ value of any edge changes during an iteration. Two additional sets, $N$ and $L$, are used as local storage: $N$ stores the edges that, together with $e$, form a triangle, and for each such triangle $(e,e',e'')$, the value $\rho=\min\{h(e'),h(e'')\}$ is inserted into $L$. Finally, $H$ denotes the $h$-index of the multi-set $L$.

From the figure, the initial supports are as follows: $h(BD)=3$, $h(AB)=h(AD)=h(AE)=h(BE)=h(DE)=2$, $h(BC)=h(CD)=1$, and $h(CF)=0$. Every edge is initially scheduled, i.e., $scheduled=\text{TRUE}$. Since $updated=\text{TRUE}$, the algorithm enters the while loop (Lines 6--31). For each scheduled edge (Lines 8--30), the algorithm constructs $N$ and $L$. We illustrate this process for two edges, $AE$ and $BD$.

\emph{Edge $AE$.} In the first triangle containing $AE$, namely $(AE,AB,BE)$, we have $N=\{AB,BE\}$ and $\rho=\min\{h(AB),h(BE)\}=\min\{2,2\}=2$, giving $L=\{2\}$. For the triangle $(AE,AD,DE)$, we update $N=\{AB,BE,AD,DE\}$ and add $\rho=\min\{h(AD),h(DE)\}=\min\{2,2\}=2$, resulting in $L=\{2,2\}$. Thus, after all triangles containing $AE$ are processed, we compute $H(AE)=h\text{-index}(L)=2$. In parallel, the edge $BD$ is also processed.

\emph{Edge $BD$.} For the first triangle $(BD,BC,CD)$, we have $N=\{BC,CD\}$ and $\rho=\min\{h(BC),h(CD)\}=\min\{1,1\}=1$, giving $L=\{1\}$. For the second triangle $(BD,AB,AD)$, we update $N=\{BC,CD,AB,AD\}$ and add $\rho=\min\{h(AB),h(AD)\}=\min\{2,2\}=2$, resulting in $L=\{1,2\}$. For the third triangle $(BD,BE,DE)$, we update $N=\{BC,CD,AB,AD,BE,DE\}$ and add $\rho=\min\{h(BE),h(DE)\}=\min\{2,2\}=2$, giving $L=\{1,2,2\}$. Hence, $H(BD)=h\text{-index}(L)=2$, decreasing from its initial value of $3$.

Since $H(BD)\neq3$, edge $BD$ satisfies the update condition. Before updating its $h$-value, the condition $H(BD)<h(e_N)\leq h(BD)$ is checked for each neighboring edge $e_N\in N(BD)$. Since none of these edges satisfies the condition, no neighboring edge is scheduled. The $h$-value of $BD$ is then updated to $2$. In the next iteration of the while loop (Lines 6--31), no further updates occur, and the loop terminates.

Finally, the trussness of each edge is set to $h(e)+2$
(Line~31-33), yielding
$trussness(E)=\{AB:4,\; AD:4,\; AE:4,\; BC:3,\; BD:4,\; BE:4,\; CD:3,\; CF:2,\; DE:4\}$.

Thus, the $3$-truss is given by the union $\bigcup\{AB,AD,AE,BC,BD,BE,CD,DE\}$, and the $4$-truss is $\bigcup\{AB,AD,AE,BD,BE,DE\}$, as shown in Fig.~\ref{fig:truss_dec}.

However, the mobile-agent framework introduces additional challenges. The agents do not know the global topology and can communicate only when they are co-located, while their movements are synchronous but not centrally controlled. We adapt the parallel truss decomposition algorithm of~\cite{WGST18} to the mobile-agent model. Each of the $n$ agents $r_i$ maintains the following variables:
\begin{enumerate}

    \item $r_i.ID$ - stores the ID of agent $r_i$. The agents have distinct IDs in the range $[0,\lambda]$, where $\lambda$ denotes the maximum agent ID. Hence, $r_i.ID$ requires $O(\log\lambda)$ bits.

    \item $r_i.nbd\_list$ - stores the IDs of the neighbors of $r_i$ and the corresponding port information. Since an agent has at most $\Delta$ neighbors, this list requires $O(\Delta\log\lambda)$ bits.

    \item $r_i.edge\_set$ - stores the edges owned by $r_i$. An edge $(r_i,r_j)$ is owned by $r_i$ only when $r_i.ID<r_j.ID$. Thus, every edge is owned by exactly one of its endpoint agents. Since an agent owns at most $\Delta$ edges, $edge\_set$ requires $O(\Delta\log\lambda)$ bits.

    \item $h(e)$ - stores the current $h$-value of every edge $e\in r_i.edge\_set$. Initially, $h(e)$ is set to the support of $e$. At termination, the trussness of $e$ is $h(e)+2$.
    
    \item $h_{\mathrm{old}}(e)$ and $h_{\mathrm{new}}(e)$ - store, respectively, the value of $h(e)$ at the beginning of an iteration and the value computed during that iteration. This ensures that all edges in one iteration are updated using the same set of $h$-values.

    \item $N(e)$ - stores the triangle-incidence information associated with each owned edge $e$. For every triangle $(e,e',e'')$ containing $e$, it stores the pair $(e',e'')$, together with the identities of their owner agents. These records are constructed during Phase~2 and retained throughout Phase~3.

    \item $L(e)$ - a multi-set associated with each owned edge $e$. During an iteration, for every stored pair $(e',e'')\in N(e)$, it stores $\min\{h_{\mathrm{old}}(e'),h_{\mathrm{old}}(e'')\}$. The multi-set is cleared and reconstructed at the beginning of every iteration.

    \item $scheduled(e)$ - indicates whether an owned edge $e$ must be processed in the current iteration. Initially, every owned edge is scheduled.
    
    \item $next\_scheduled(e)$ - indicates whether an edge must be processed in the next iteration. It is reset to $false$ at the beginning of every iteration and becomes $true$ if the agent receives a scheduling request for that edge.
    
    \item $r_i.change$ - is set to $1$ when none of the edges owned by $r_i$ is scheduled for the next iteration, and to $0$ otherwise.

\end{enumerate}

\paragraph*{\textbf{Phase 1 (\textsc{Protocol MYN})}}
Initially, each node of $G$ hosts one agent. Every agent executes \textsc{Protocol MYN} to meet all its neighbors and record their IDs and the corresponding port numbers, as described in Phase~1 of Section~\ref{sec: traingle_counting}. Each edge is assigned to exactly one of its endpoint agents. In particular, $r_i$ stores an incident edge $(r_i,r_j)$ in $r_i.edge\_set$ only if $r_i.ID<r_j.ID$. We say that $r_i$ is the \emph{owner} of this edge. Phase~1 takes $O(\Delta\log\lambda)$ rounds.

\paragraph*{\textbf{Phase 2 (Initialization)}}
The agents now compute the support of every owned edge. When two adjacent agents $r_i$ and $r_j$ meet, they exchange their neighbor lists and find their common neighbors. If $r_i.ID<r_j.ID$, then $r_i$ owns the edge $e=(r_i,r_j)$. The number of common neighbors of $r_i$ and $r_j$ is recorded as $support(e)$, and $h(e)$ is initialized to this value. The triangle information required in Phase~3 is also stored during this phase. For every common neighbor $r_p$, the edge $e=(r_i,r_j)$ belongs to the triangle formed by $r_i$, $r_j$, and $r_p$. Let $e'=(r_i,r_p)$ and $e''=(r_j,r_p)$ be the other two edges of this triangle. The owner of each edge is determined directly from the IDs of its endpoints. Agent $r_i$ stores the pair $(e',e'')$, together with the owners of these edges, in $N(e)$. Thus, $N(e)$ records all triangles containing $e$. It is constructed once in Phase~2 and retained throughout the algorithm; triangle incidences are not recomputed in Phase~3. Phase~2 takes $O(\Delta\log\lambda)$ rounds and uses $O(\Delta^2\log\lambda)$ bits.

\paragraph*{\textbf{Phase 3 ($h$-Value Updates)}}
The agents repeatedly update the $h$-values until no edge needs further processing. Each iteration consists of two executions of \textsc{Protocol MYN}: the first collects all required $h$-values, and the second delivers all scheduling requests.

At the beginning of an iteration, every agent sets $h_{\mathrm{old}}(e)\leftarrow h(e)$ and $h_{\mathrm{new}}(e)\leftarrow h_{\mathrm{old}}(e)$ for every edge $e$ it owns, and initializes $next\_scheduled(e)\leftarrow false$. All updates in the iteration are computed using the fixed values $h_{\mathrm{old}}$.

\begin{enumerate}
    \item \textbf{Identify scheduled edges.}
    Each agent identifies the edges $e\in edge\_set$ for which $scheduled(e)=true$. An agent with no scheduled edge performs no $h$-index computation, but remains active because other agents may require its stored $h$-values or may send it scheduling requests.

    \item \textbf{Collect the required $h$-values.}
    All agents execute \textsc{Protocol MYN} once. Whenever two agents meet, they exchange the identifiers and $h_{\mathrm{old}}$-values of all the edges they own. Each agent stores the received records in a temporary table indexed by edge identifiers. To see why one execution is sufficient, consider an edge $e=(r_i,r_j)$ owned by $r_i$ and a triangle containing a common neighbor $r_p$. The other two edges are $e'=(r_i,r_p)$ and $e''=(r_j,r_p)$. The owners of $e'$ and $e''$ are among $r_i$, $r_j$, and $r_p$. Since both $r_j$ and $r_p$ are neighbors of $r_i$, agent $r_i$ obtains the required values $h_{\mathrm{old}}(e')$ and $h_{\mathrm{old}}(e'')$ while meeting all its neighbors. Therefore, the same execution of \textsc{Protocol MYN} supplies the information required for all scheduled edges owned by $r_i$. Under our communication model, co-located agents can exchange all these stored records in one round.

    \item \textbf{Compute the new $h$-values locally.}
    For every scheduled edge $e$, its owner clears $L(e)$ and constructs $L(e)=\{\min\{h_{\mathrm{old}}(e'),h_{\mathrm{old}}(e'')\}:(e',e'')\in N(e)\}$. It then computes $H(e)$, the $h$-index of $L(e)$. If $H(e)<h_{\mathrm{old}}(e)$, it sets $h_{\mathrm{new}}(e)\leftarrow H(e)$. For every edge $e_N$ appearing in $N(e)$ that satisfies $H(e)<h_{\mathrm{old}}(e_N)\leq h_{\mathrm{old}}(e)$, the agent stores a scheduling request for the owner of $e_N$. These computations use only locally stored information and therefore require no additional communication rounds.

    \item \textbf{Deliver scheduling requests.}
    All agents execute \textsc{Protocol MYN} a second time and deliver all stored scheduling requests when they meet the corresponding edge owners. One execution is sufficient because the owner of every edge appearing in $N(e)$ is either the current agent or one of its immediate neighbors. When an agent receives a request for an edge $e$ that it owns, it sets $next\_scheduled(e)\leftarrow true$.

    \item \textbf{Complete the iteration.}
    Every agent updates each owned edge by setting $h(e)\leftarrow h_{\mathrm{new}}(e)$ and $scheduled(e)\leftarrow next\_scheduled(e)$. It sets $r_i.change=1$ if none of its owned edges is scheduled, and sets $r_i.change=0$ otherwise.
    
    \item\label{termination} \textbf{Checking for Termination.}
    The $change$ values are aggregated at the leader through the BFS tree, and their binary product is broadcast to all agents in $O(h_T)$ rounds by Lemma~\ref{spanning_tree_communication}. If the product is $1$, no edge remains scheduled, and all agents terminate. Otherwise, another iteration of Phase~3 begins.

\end{enumerate}

\begin{lem}\label{lem:truss_phases}
Phases~1 and~2 each take $O(\Delta\log\lambda)$ rounds, while one iteration of Phase~3 takes $O(\Delta\log\lambda+h_T)$ rounds.
\end{lem}

\begin{proof}
By Lemma~\ref{lemma: list_neighbours}, one execution of \textsc{Protocol MYN} allows every agent to meet all its neighbors in $O(\Delta\log\lambda)$ rounds. Hence, Phase~1 records the neighbor, port, and edge-ownership information within this time. In Phase~2, the agents execute \textsc{Protocol MYN} again, exchange their neighbor lists, and simultaneously compute the support and store the triangle-incidence list $N(e)$ for every owned edge, giving the same bound. In each iteration of Phase~3, one execution of \textsc{Protocol MYN} collects the $h_{\mathrm{old}}$-values required for all scheduled owned edges, after which the agents construct $L(e)$ and compute the new $h$-values locally. A second execution delivers all scheduling requests to the corresponding edge owners. Since co-located agents can exchange all their stored information in one round, these operations require $O(\Delta\log\lambda)$ rounds in total, rather than one execution for each edge. Finally, termination detection through the BFS tree takes $O(h_T)$ rounds. Therefore, one iteration of Phase~3 takes $O(\Delta\log\lambda+h_T)$ rounds.
\end{proof}

\begin{lem}\label{lem:truss_memory}
The truss decomposition algorithm uses $O(\Delta^2\log\lambda)$ bits of memory per agent.
\end{lem}

\begin{proof}
For the agent ID, port number, edge identifier, or $h$-values, an agent needs $O(\log \lambda)$ bits. The neighbor list, owned-edge list, and the values $h_{\mathrm{old}}$, $h_{\mathrm{new}}$, $scheduled$, and $next\_scheduled$ for at most $\Delta$ owned edges require $O(\Delta\log\lambda)$ bits. For each owned edge $e$, the list $N(e)$ contains at most $\Delta-1$ triangle records, while $L(e)$ contains at most $\Delta-1$ $h$-values; thus, each requires $O(\Delta\log\lambda)$ bits. Since an agent owns at most $\Delta$ edges, all such lists require $O(\Delta^2\log\lambda)$ bits. During an iteration, an agent may additionally receive the identifiers and $h_{\mathrm{old}}$-values of at most $\Delta$ edges from each of its at most $\Delta$ neighbors and may store $O(\Delta^2)$ scheduling requests, which also require $O(\Delta^2\log\lambda)$ bits. Therefore, the total memory usage is $O(\Delta^2\log\lambda)$ bits per agent.
\end{proof}

\begin{thm}\label{thm:truss_decompose}
Let $G$ be an arbitrary, simple, connected graph with $n$ nodes, $m$ edges, maximum degree $\Delta$, and diameter $D$. Suppose $n$ mobile agents with distinct IDs from $[0,\lambda]$, are initially placed arbitrarily in $G$, and let $\lambda$ be the maximum agent ID. After initialization, the \textsc{Truss Decomposition Problem} can be solved in $O(m\Delta\log\lambda+mD)$ rounds using $O(\Delta^2\log\lambda)$ bits of memory per agent.
\end{thm}

\begin{proof}
By Lemma~\ref{lem:truss_phases}, Phases~1 and~2 each take $O(\Delta\log\lambda)$ rounds. In every iteration of Phase~3, one execution of \textsc{Protocol MYN} collects the $h_{\mathrm{old}}$-values required for all scheduled owned edges, the new $h$-values are computed locally using the stored lists $N(e)$, and a second execution delivers all scheduling requests. Termination detection takes $O(h_T)$ rounds. Thus, one iteration takes $O(\Delta\log\lambda+h_T)$ rounds. The monotone $h$-index process of~\cite{WGST18} converges in at most $m$ iterations, giving $O(m\Delta\log\lambda+mh_T)=O(m\Delta\log\lambda+mD)$ rounds by Remark~\ref{rem:D-h_t}. For correctness, at convergence let $E_t=\{e:h(e)\geq t\}$. The $h$-index condition implies that every edge in $E_t$ belongs to at least $t$ triangles whose other edges also belong to $E_t$; hence, $E_t$ forms a $(t+2)$-truss. Conversely, an edge belonging to a $(t+2)$-truss never decreases below $t$, since it always has at least $t$ supporting triangles within that truss. Therefore, $h(e)=trussness(e)-2$, and the output $h(e)+2$ is correct for every edge. The memory bound follows from Lemma~\ref{lem:truss_memory}.
\end{proof}

\begin{remark}
The $O(\Delta^2\log\lambda)$ memory allows an agent to retain the triangle-incidence lists $N(e)$ and the $h$-values and scheduling records required for all its owned edges. Consequently, one execution of \textsc{Protocol MYN} collects all required values and another delivers all scheduling requests, preserving the $O(\Delta\log\lambda+h_T)$ cost per iteration. With only $O(\Delta\log\lambda)$ memory, an agent may retain the information for only one owned edge at a time and may need to repeat the communication for up to $\Delta$ edges, introducing an additional multiplicative factor of $\Delta$ in the round complexity.
\end{remark}

%% file: application.tex
\subsection{Triangle Centrality}
The concept of \emph{Triangle Centrality} was introduced in ~\cite{B2021} by Paul Burkhardt. The concept may be useful for finding important vertices in a graph based on the concentration of triangles surrounding each vertex. In this section, we employ $n$ mobile agents, each with a distinct ID, starting at each distinct node of an arbitrarily connected anonymous graph, to compute the triangle centrality for each node. Mathematically, Triangle Centrality, $TC(v)$ of a node $v\in G$ is formulated in ~\cite{B2021} as: 
$$TC(v)=\frac{\frac{1}{3}\sum_{u\in N^+_{\mathbf{T}}(v)}\mathbf{T}(u)+\sum_{w\in N(v)\setminus N_{\mathbf{T}}(v)}\mathbf{T}(w)}{T(G)}$$, where, $N(v)$ is the neighborhood set of $v$, $N_{\mathbf{T}}(v)$ is the set of neighbors that are in triangles with $v$, and $N^+_{\mathbf{T}}(v)$ is the closed set that includes $v$. $\mathbf{T}(v)$ and $\mathbf{T}(G)$ denote the respective triangle counts based on $v$ and total triangle count in $G$. Mathematically, 
$N(v)=\{u:(u,v)\in E(G)\}$, $N_{\mathbf{T}}(v)=\{u\in N(v):N(u)\cap N(v)\neq\phi\}$ and $N^+_{\mathbf{T}}(v)=\{v\}\cup N_{\mathbf{T}}(v)$.
Now, we outline our algorithm that computes $TC(v)$ for a node $v\in G$ using the mobile agents that run in these $3$ phases: 
\begin{enumerate}
    \item \textbf{Computing $T(v)$ and $T(G)$} -- Using the triangle-counting algorithm described in Section~\ref{sec: traingle_counting}, each agent computes $T(v)$ in $O(\Delta\log\lambda)$ rounds. If $T(G)$ is unknown, the local triangle counts are aggregated and disseminated through the spanning tree in an additional $O(n)$ rounds (Theorem~\ref{thm:gen_config}($1$)). Thus, this phase takes $O(\Delta\log\lambda)$ rounds when $T(G)$ is known and $O(n+\Delta\log\lambda)$ rounds otherwise.

   \item \textbf{Computing $N(v)$, $N^+_{\mathbf{T}}(v)$ and      $N_{\mathbf{T}}(v)$} -
    The agents record their neighborhoods $N(v)$ in $O(\Delta\log\lambda)$ rounds (Lemma~\ref{lemma: list_neighbours}). In another $O(\Delta\log\lambda)$ rounds, each agent meets its neighbors again and exchanges neighborhood lists and the triangle counts computed in the preceding step. For each $u\in N(v)$, the agent at $v$ includes $u$ in $N_{\mathbf{T}}(v)$ if $N(u)\cap N(v)\neq\emptyset$, and constructs $N^+_{\mathbf{T}}(v)=N_{\mathbf{T}}(v)\cup\{v\}$. During these meetings, counting each neighbor once, it computes $s=\mathbf{T}(v)+\sum_{u\in N_{\mathbf{T}}(v)}\mathbf{T}(u)=\sum_{u\in N^+_{\mathbf{T}}(v)}\mathbf{T}(u)$ and $x=\sum_{u\in N(v)}\mathbf{T}(u)$. It then obtains $\sum_{w\in N(v)\setminus N_{\mathbf{T}}(v)}\mathbf{T}(w)=x-s+\mathbf{T}(v)$. This step takes $O(\Delta\log\lambda)$ rounds.

    \item \textbf{Computing $TC(v)$} -
    Having obtained these sums and $\mathbf{T}(G)$, the agent at $v$ evaluates $TC(v)=\frac{\frac{1}{3}\sum_{u\in N^+_{\mathbf{T}}(v)}\mathbf{T}(u)+\sum_{w\in N(v)\setminus N_{\mathbf{T}}(v)}\mathbf{T}(w)}{\mathbf{T}(G)}$. Here, we assume $\mathbf{T}(G)>0$; if $\mathbf{T}(G)=0$, we define $TC(v)=0$ for every node $v$.
\end{enumerate}
To this end, we have the following theorem.
\begin{thm}
    Let $G$ be an $ n$-node arbitrary, simple connected graph with a maximum degree $\Delta$. Let $n$ mobile agents with distinct IDs in the range $[0,n^c]$ with the highest ID $\lambda$, where $c$ is an arbitrary constant, be placed at each $n$ node of $G$ in an initial dispersed configuration. Then, the \emph{triangle centrality} of each node $v\in G$ can be calculated in $O(\Delta\log\lambda)$ rounds if $\mathbf{T(G)}$ is known and in $O(n+\Delta\log\lambda)$ rounds, if  $\mathbf{T}(G)$ is unknown. $\mathbf{T}(G)$ is the total triangle count of the graph $G$. 
\end{thm}

\subsection{Local Clustering Coefficient}
The local clustering coefficient of a node in a graph is used to quantify how close its neighbors are to being a clique (complete graph), i.e., how well connected the network is around a particular node. Mathematically, the local clustering coefficient ($LCC$) of a node $v\in G$ can be written as $LCC(v)=\frac{2T(v)}{\delta(v)(\delta(v)-1)}$,(from ~\cite{SU23}).For $\delta(v)<2$, we define $LCC(v)=0$. Here $\delta(v)$ denotes the degree of the node $v$. This metric can be easily calculated using mobile agents. $T(v)$ for a node $v$ can be calculated in $O(\Delta\log\lambda)$ rounds [Lemma~\ref{lemma: local_triangle_counting}]. The agent that already knows $\delta(v)$ at $v$ can now evaluate $LCC(v)$ using the formula. 

\begin{thm}
    Let $G$ be an $n$-node, arbitrary, simple connected graph with a maximum degree $\Delta$. Let $n$ mobile agents with distinct IDs in the range $[0,n^c]$ with the highest ID $\lambda$, where $c$ is an arbitrary constant, be placed at each $n$ node of $G$ in an initial dispersed configuration. Then, the \emph{Local Clustering Coefficient} of each node $v\in G$, $LCC(v)$ can be calculated in $O(\Delta\log\lambda)$ rounds. 
\end{thm}

%% file: simulation.tex
\section{Simulations}\label{sec:experiments}

In this section, we evaluate the performance of the proposed $h$-index-based truss decomposition algorithm. We implemented Algorithm~\ref{alg:k-truss-parallel} in \textbf{Python}, and all experimental evaluations were carried out in the \textbf{Google Colab} environment, using its cloud-based computational infrastructure to execute the iterative upgradation phases. The code and details can be found at~\cite{sim_url}. We evaluate the performance of our algorithm using two distinct classes of synthetic graph models, representing different topological properties:

\begin{enumerate}
    \item \textbf{Erd\H{o}s--R\'enyi (ER) Random Graphs:} We utilize the $G(n, p)$ model to generate baseline random networks. In this model, $n$ nodes are connected independently with a fixed probability $p$. These graphs allow us to study the algorithm's behavior across a wide range of edge densities and to evaluate how the number of truss-upgradation phases scales with the total number of edges $m$.
    
    \item \textbf{Small-World Graphs (Watts--Strogatz Model):} To simulate more realistic network structures, we employ the Watts--Strogatz model. These graphs exhibit high local clustering and short average path lengths. Starting from a regular ring lattice where each node is connected to its $k$ nearest neighbors, edges are rewired with a probability $\beta$. This dataset is particularly significant for truss decomposition as it contains the dense, localized triangle structures typically found in social and biological networks.
\end{enumerate}

In our experiments, we measured the actual number of truss-decomposition phases required in practice compared to the worst-case theoretical bound, since the running time depends heavily on the graph topology. Although the theoretical analysis allows up to $m$ phases, where $m$ is the number of edges in the graph, the algorithm may converge much faster in practice when many edges are updated or removed simultaneously during a phase. Thus, the experiments help quantify the gap between the guaranteed upper bound and the observed running behavior.

\subsection{Effect of the number of edges (ER Random Graphs)}

\begin{figure}[t]
    \centering
    \includegraphics[width=0.8\linewidth]{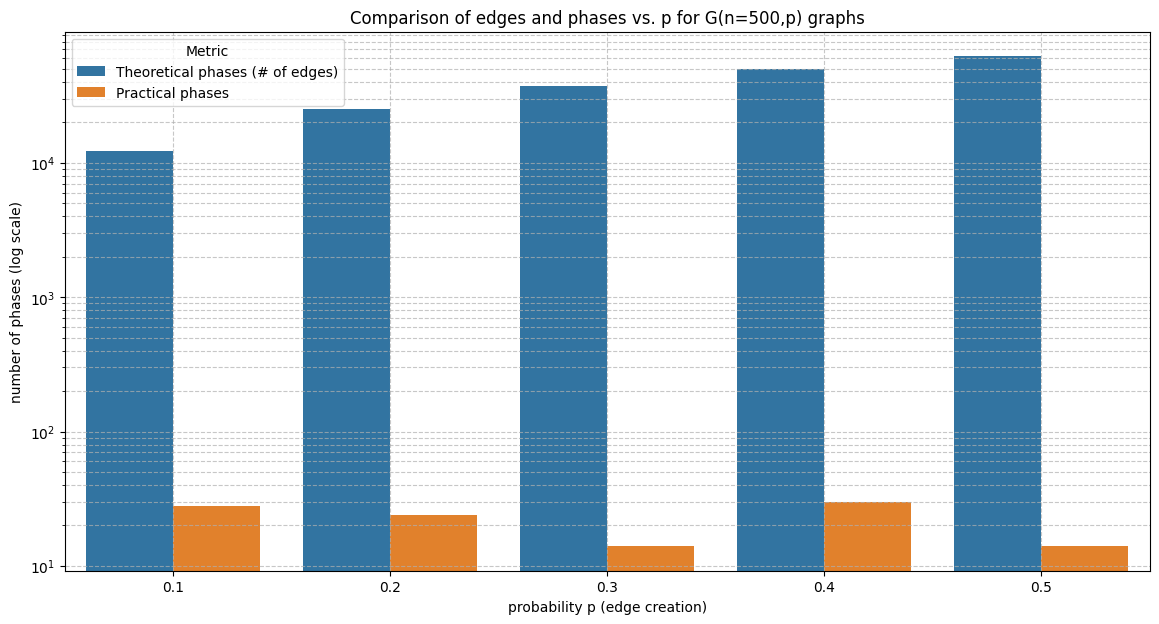}
    \caption{Comparison between the number of edges and the practical number
    of truss-upgradation phases for random graphs $G(n=500,p)$, with varying
    edge-creation probability $p$ ($x$-axis). The $y$-axis (number of phases) is shown on a logarithmic scale.}
    \label{fig:truss-p}
\end{figure}

We first fix $n=500$ and vary the edge-creation probability $p$. The result is shown in Fig.~\ref{fig:truss-p}. The blue bars denote the number of edges in the generated graph, which gives the theoretical worst-case upper bound on the number of upgradation phases. The orange bars denote the actual number of phases required by the $h$-index-based truss decomposition algorithm. The figure shows that, although the number of edges increases rapidly as $p$ grows, the practical number of phases remains small in comparison. In all tested cases, the algorithm converges within only a few tens of phases, while the number of edges is in the order of tens of thousands. This indicates that many edges update their values simultaneously, rather than progressing one by one.

\subsection{Effect of the number of nodes (ER Random Graphs)}

\begin{figure}[t]
    \centering
    \includegraphics[width=0.8\linewidth]{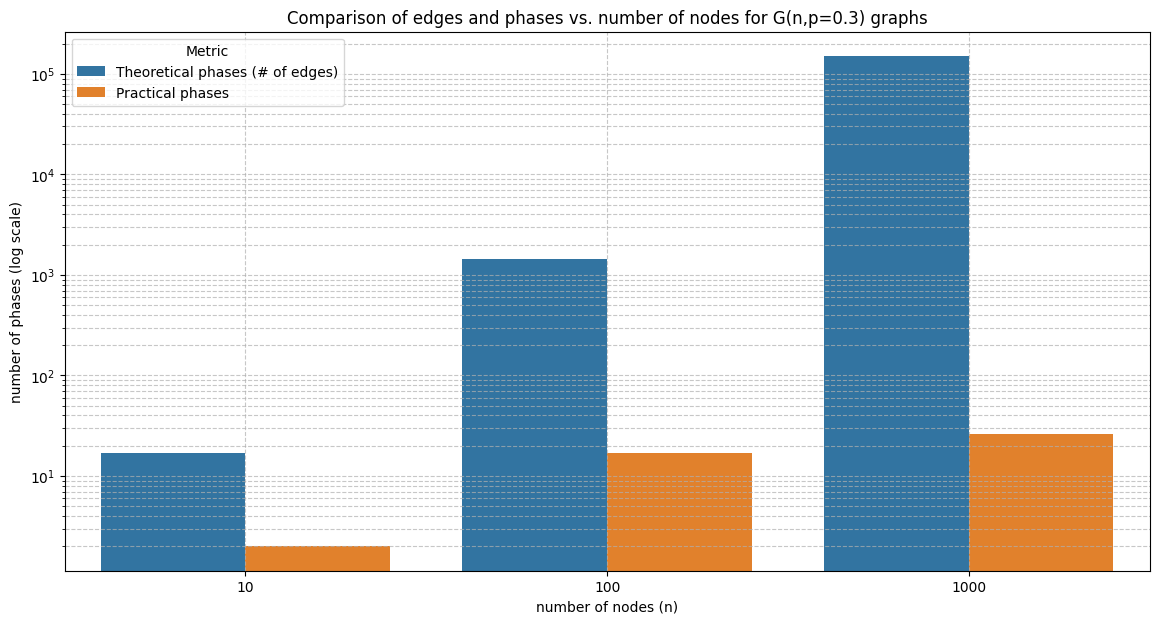}
    \caption{Comparison between the number of edges and the practical number
    of truss-upgradation phases for random graphs $G(n,p=0.3)$, with varying
    number of nodes $n$ ($x$-axis). The $y$-axis (number of phases) is shown in logarithmic scale.}
    \label{fig:truss-n}
\end{figure}

We next fix the edge probability to $p=0.3$ and vary the number of nodes $n$. The result is shown in Fig.~\ref{fig:truss-n}. The blue bars denote the number of edges in the generated graph, which gives the theoretical worst-case upper bound on the number of upgradation phases, while the orange bars denote the actual number of phases required by the $h$-index-based truss decomposition algorithm. The figure shows that, although the number of edges increases sharply as $n$ grows, the practical number of phases increases much more slowly and remains small in comparison. Even for the largest graph, the observed number of phases is far below the theoretical bound. This shows that the growth in graph size does not necessarily translate into a proportional increase in the number of upgradation phases.

\subsection{Effect of local neighborhood size in small-world graphs}

\begin{figure}[t]
    \centering
    \includegraphics[width=0.95\linewidth]{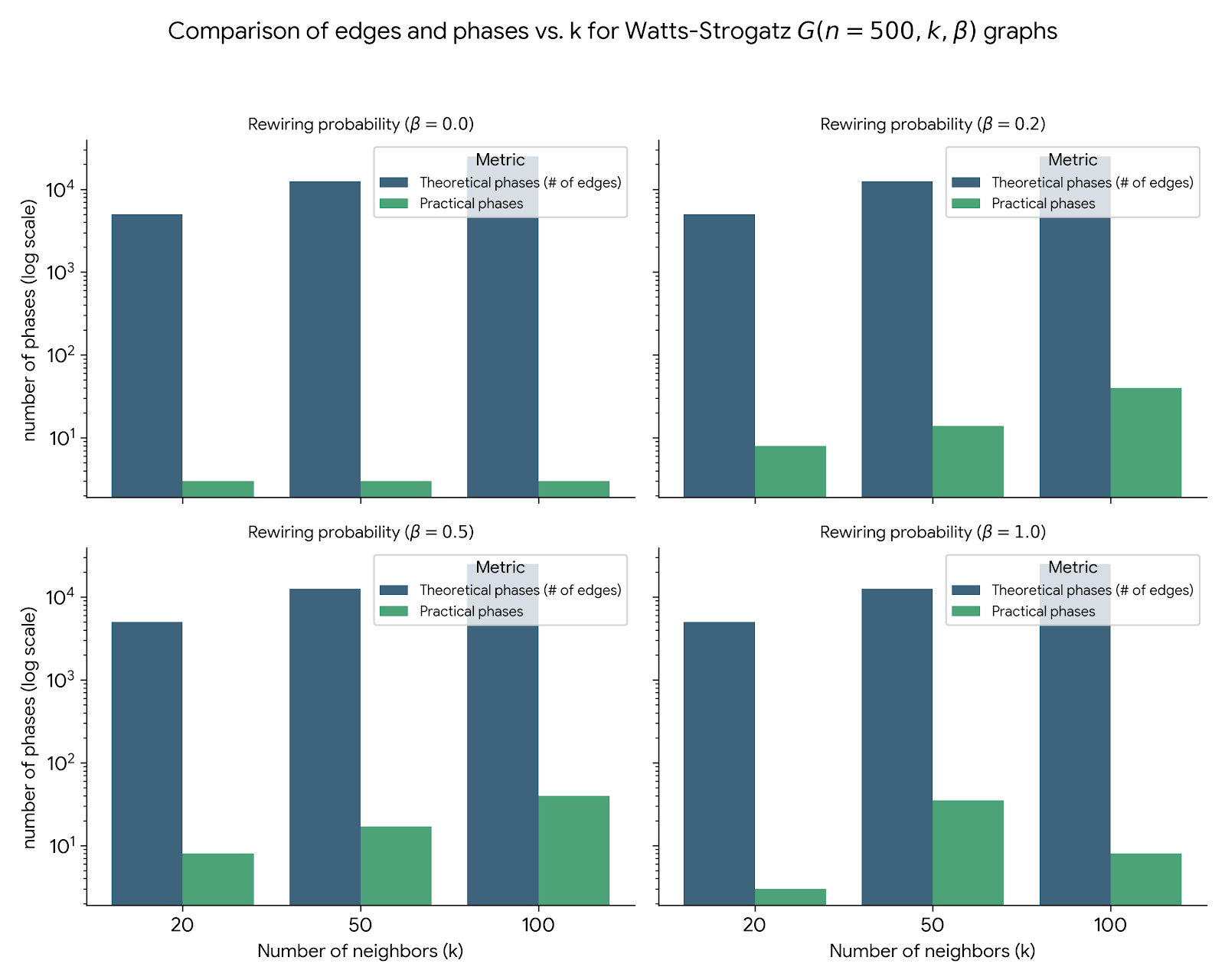}
    \caption{Comparison between the number of edges and the practical number
    of truss-upgradation phases for Watts--Strogatz small-world graphs with
    $n=500$, varying neighborhood size $k$ ($x$-axis), and rewiring probability $\beta$.
    The $y$-axis (number of phases) is shown on a logarithmic scale.}
    \label{fig:truss-ws}
\end{figure}

We also evaluate the algorithm on Watts--Strogatz small-world graphs with $n=500$. Here, we vary the neighborhood size $k$ and the rewiring probability $\beta$, as shown in Fig.~\ref{fig:truss-ws}. The blue bars denote the number of edges in the graph, which gives the theoretical worst-case upper bound on the number of upgradation phases, while the green bars denote the actual number of phases required by the $h$-index-based truss decomposition algorithm. The figure shows that, for all tested values of $k$ and $\beta$, the practical number of phases remains much smaller than the theoretical bound. Even when the number of edges becomes large, the algorithm converges in only a few phases or a few tens of phases. This suggests that the clustered local structure of small-world graphs allows many edge values to stabilize together.

\subsection{Discussion and Summary of Experimental Findings}

The experiments show that the proposed mobile-agent-based truss decomposition algorithm performs efficiently on both Erd\H{o}s--R'enyi random graphs and Watts--Strogatz small-world graphs. In every tested instance, the number of upgradation phases is much smaller than the theoretical upper bound of $m$, mainly because many edges update and stabilize simultaneously when their local triangle structures are similar. Since the costs of \textsc{Protocol MYN} and global communication per phase remain fixed, this reduction in the number of phases is the main source of practical improvement. The $m$-phase worst-case upper bound remains valid for arbitrary anonymous graphs, including adversarial cases with limited progress per phase. Hence, although the worst-case round complexity remains $O(\Delta\log\lambda)+O(\Delta\log\lambda)+mO(\Delta\log\lambda+D)$, the tested graph families converge much faster in practice. Thus, the simulations complement the theoretical analysis by showing favorable empirical performance alongside the worst-case guarantees."

%% file: conclusion.tex
\section{Concluding Remarks and Future Work}\label{sec:conclusion}

In this paper, we presented mobile-agent-based algorithms for triangle counting in arbitrary anonymous graphs and extended them to truss decomposition, triangle centrality, and local clustering coefficient computation. The framework demonstrates how limited-memory agents with only local communication can perform non-trivial graph analytics in anonymous networks. We also provided simulation-based evaluations of the truss decomposition algorithm and compared the observed number of truss-upgradation phases with the theoretical bounds. An important future direction is to study these problems using fewer agents. At one extreme, a single agent can solve them through complete map construction, but exploration in anonymous graphs is highly expensive. As shown in~\cite{map_const_opodis10}, without prior knowledge of $n$, exploration requires $O(n^6\Delta^2\log n)$ rounds and $O(n^3\Delta^2\log n\log\Delta)$ bits of memory, whereas with a marked node, equivalently two agents with one fixed marker, it requires $O(n^3\Delta)$ rounds and $O(n\Delta\log n)$ bits. The same work also shows strong time--memory trade-offs, where reducing memory may cause exponential time overhead. Increasing the number of agents does not necessarily remove communication bottlenecks, since the improvement depends on the graph topology and initial placement. For example, in a chain of triangles connected by single edges, placing one agent in each triangle may still prevent agents from meeting through a single-hop movement. Thus, even with $\Theta(n)$ agents, substantial exploration may be required to establish communication. It would therefore be interesting to determine whether agents placed $t$ hops apart inherently require $O(\Delta^t\log\lambda)$ rounds to communicate. Understanding the trade-offs among the number of agents, memory, communication cost, initial placement and knowledge, and graph structure remains an important open problem. Another useful direction is to study triangle-based analytics in the presence of faulty or crash-prone agents.